\documentclass[twocolumn, hyperref=colorlinks,linkcolor=blue,citecolor=blue,urlcolor=red]{aastex}

\usepackage{threeparttable}
\usepackage{graphicx}
\usepackage{placeins}
\usepackage{float}
\usepackage{url}
\usepackage{natbib}
\usepackage{xspace}  
\usepackage{amssymb}
\usepackage{amsmath} 

\def\gs{\mathrel{\raise1.16pt\hbox{$>$}\kern-7.0pt
\lower3.06pt\hbox{{$\scriptstyle \sim$}}}}
\def\ls{\mathrel{\raise1.16pt\hbox{$<$}\kern-7.0pt
\lower3.06pt\hbox{{$\scriptstyle \sim$}}}}
\def\gtsima{\, {\buildrel > \over \sim} \,}
\def\ltsima{\, {\buildrel < \over \sim} \,}
\def\prosima{\, {\buildrel \propto \over \sim} \,}
\def\gsim{\lower.5ex\hbox{\gtsima}}
\def\lsim{\lower.5ex\hbox{\ltsima}}
\def\simgt{\lower.5ex\hbox{\gtsima}}
\def\simlt{\lower.5ex\hbox{\ltsima}}
\def\simpr{\lower.5ex\hbox{\prosima}}

\begin{document}

\title{Connecting galaxy structure, quenching, and halo environments during cosmic noon}

\author[0000-0002-9272-5978]{Ziwen Zhang}
\affiliation{Department of Astronomy, University of Science and Technology of China, Hefei, Anhui 230026, China; Email: ziwen@mail.ustc.edu.cn, whywang@ustc.edu.cn, enciwang@ustc.edu.cn}
\affiliation{School of Astronomy and Space Science, University of Science and Technology of China, Hefei 230026, China}

\author[0009-0004-7042-4172]{Cheng Jia}
\affiliation{Department of Astronomy, University of Science and Technology of China, Hefei, Anhui 230026, China; Email: ziwen@mail.ustc.edu.cn, whywang@ustc.edu.cn, enciwang@ustc.edu.cn}
\affiliation{School of Astronomy and Space Science, University of Science and Technology of China, Hefei 230026, China}

\author[0000-0002-4911-6990]{Huiyuan Wang}
\affiliation{Department of Astronomy, University of Science and Technology of China, Hefei, Anhui 230026, China; Email: ziwen@mail.ustc.edu.cn, whywang@ustc.edu.cn, enciwang@ustc.edu.cn}
\affiliation{School of Astronomy and Space Science, University of Science and Technology of China, Hefei 230026, China}

\author[0000-0003-1588-9394]{Enci Wang}
\affiliation{Department of Astronomy, University of Science and Technology of China, Hefei, Anhui 230026, China; Email: ziwen@mail.ustc.edu.cn, whywang@ustc.edu.cn, enciwang@ustc.edu.cn}
\affiliation{School of Astronomy and Space Science, University of Science and Technology of China, Hefei 230026, China}

\author[0000-0002-4597-5798]{Yangyao Chen}
\affiliation{School of Astronomy and Space Science, Nanjing University, Nanjing, Jiangsu 210093, China}
\affiliation{Key Laboratory of Modern Astronomy and Astrophysics, Nanjing University, Ministry of Education, Nanjing, Jiangsu 210093, China}

\author[0000-0001-5356-2419]{Houjun Mo}
\affiliation{Department of Astronomy, University of Massachusetts, Amherst, MA 01003-9305, USA}
  
\author[0009-0006-0435-9469]{Xiong Luo}
\affiliation{Department of Astronomy, University of Science and Technology of China, Hefei, Anhui 230026, China; Email: ziwen@mail.ustc.edu.cn, whywang@ustc.edu.cn, enciwang@ustc.edu.cn}
\affiliation{School of Astronomy and Space Science, University of Science and Technology of China, Hefei 230026, China}

\author[0009-0000-7307-6362]{Cheqiu Lyu}
\affiliation{Department of Astronomy, University of Science and Technology of China, Hefei, Anhui 230026, China; Email: ziwen@mail.ustc.edu.cn, whywang@ustc.edu.cn, enciwang@ustc.edu.cn}
\affiliation{School of Astronomy and Space Science, University of Science and Technology of China, Hefei 230026, China}

\author[0000-0002-0846-7591]{Jie Song}
\affiliation{Department of Astronomy, University of Science and Technology of China, Hefei, Anhui 230026, China; Email: ziwen@mail.ustc.edu.cn, whywang@ustc.edu.cn, enciwang@ustc.edu.cn}
\affiliation{School of Astronomy and Space Science, University of Science and Technology of China, Hefei 230026, China}

\begin{abstract}
The connection between galaxy structure, star formation activity, and large-scale environments provides important insights into the evolution of massive galaxies. However, the relative importance of different galaxy properties in predicting quenching and the environmental dependence of structurally distinct galaxy populations remain poorly constrained at cosmic noon ($1\leq z<3$). In this work, we investigate massive galaxies with $\log M_*/{\rm M}_{\odot}>10.5$ at $1\leq z<3$ using deep JWST/NIRCam imaging from the COSMOS-Web survey, combining measurements of central stellar mass surface density ($\Sigma_1$), star formation activity, and galaxy clustering. We find that $\Sigma_1$ is the most informative predictor of quenching among the considered properties, including stellar mass, effective radius, formation redshift, and local overdensity. The quenched fraction shows a strong dependence on $\Sigma_1$, increasing rapidly above a characteristic $\Sigma_1$. We further show that compact quenched galaxies exhibit a tendency toward stronger clustering than compact and extended star-forming galaxies, while compact and extended star-forming galaxies show similar clustering properties. These results highlight the importance of $\Sigma_1$ in tracing quenching and reveal that structural diversity among star-forming galaxies is not strongly reflected in their large-scale environments, providing new constraints on the interplay between galaxy structure, quenching, and halo environments during cosmic noon.

\end{abstract}


\section{Introduction}

Observations of the nearby Universe have revealed a clear bimodality in galaxy colors and star formation activity, with galaxies occupying two distinct populations of actively star-forming and quenched systems \citep{Strateva2001,Brinchmann2004}. The transition from star-forming to quenched galaxies is particularly prominent among massive galaxies, which exhibit higher quenched fractions and provide important insights into the physical processes responsible for shutting down star formation. Deep observations have revealed that massive quenched galaxies were already in place at $z\sim2-3$ \citep{Ilbert2013}, during cosmic noon ($1\leq z < 3$) when the cosmic star formation rate density reached its peak. Understanding the physical processes responsible for quenching massive galaxies remains a central challenge in galaxy evolution.

Galaxy quenching is generally considered to result from a combination of internal and external processes. Internal mechanisms regulate star formation by altering the availability and physical state of cold gas within galaxies. Stellar and Active Galactic Nuclei (AGN) feedback can expel gas from galaxies or heat the surrounding gas, reducing the fuel available for star formation \citep{Dekel1986, Fabian-12, Zhang2021, chenyy2025, Chen2025, li2025, Lyu2026,Yu2026}, while morphological transformation can stabilize gas against gravitational collapse and suppress the formation of new stars \citep{Martig-09}. External mechanisms are associated with the surrounding environment, including ram-pressure stripping, merger and tidal interactions, which can remove or redistribute gas reservoirs and thereby suppress star formation \citep{Gunn1972, Moore1996}. Halo heating prevents efficient gas cooling and suppresses the replenishment of cold gas through accretion \citep{Dekel2006, Mo2024}. The relative importance of these mechanisms depends on galaxy properties, environment, and evolutionary stage \citep{li2025}. For massive galaxies, increasing evidence suggests that internal processes play an important role in regulating quenching, as their deep gravitational potentials make environmental processes less efficient at regulating gas \citep{Peng2010, Bluck2014}, while their structural evolution is closely linked to the shutdown of star formation \citep{fang2013, Barro2017ApJ}.

The connection between galaxy structure and quenching has been extensively investigated in the local Universe. Using SDSS observations, \citet{Kauffmann2003c} showed that stellar surface density is more strongly associated with stellar population age than stellar mass, with more compact galaxies preferentially hosting older stellar populations. Subsequent studies linked the internal distribution of stellar mass to star formation activity, showing that galaxies with prominent bulges and higher Sérsic indices are more likely to be quenched \citep{Driver2006, Bell2008}. \citet{fang2013} further quantified this relation using the central stellar mass surface density within 1 kpc ($\Sigma_1$). They showed that quenched galaxies preferentially occupy the high-$\Sigma_1$ regime, with the quenched fraction increasing rapidly above a characteristic density threshold. They further demonstrated that $\Sigma_1$ is tightly correlated with central velocity dispersion. Consistent with this picture, \citet{Hong2023} showed that central velocity dispersion is closely associated with galaxy quenching, highlighting the role of galaxy dynamical state in regulating star formation. The distribution of galaxies in the $\Sigma_1$–sSFR plane exhibits a L-shaped structure, where quenched galaxies occupy a relatively narrow high-$\Sigma_1$ sequence, whereas star-forming galaxies span a broader range of central densities, including both extended low-density and compact high-density galaxies \citep{Tacchella2016, Barro2017ApJ,Mosleh2017}. At higher redshifts, this structural connection is further supported by the existence of compact quenched galaxies at $z\sim2$, known as “red nuggets” \citep{van2010, Bruce2012}, while compact star-forming galaxies at similar epochs exhibit similarly dense and spheroid-like structures \citep{Barro2013,Barro2014}. 

These observational trends have motivated evolutionary scenarios linking different structural states along the quenching pathway. A widely discussed pathway proposes that extended star-forming galaxies undergo structural transformation through dissipative processes, including gas compaction and violent disk instabilities, which drive rapid central stellar mass growth \citep{Dekel2009, Zolotov2015, Tacchella2016, Lapiner2023, chenyy2025}. These processes can lead to the formation of compact star-forming galaxies, which subsequently quench and evolve into compact quenched galaxies \citep{Barro2017ApJ, Wang2018, wangk2026}. However, this interpretation is complicated by progenitor bias, which arises because newly quenched galaxies continuously join the quenched population over cosmic time. In this scenario, the observed evolution of quenched galaxy properties, such as the increase in size and decrease in compactness toward lower redshifts \citep{Yang2025}, may partly reflect changes in the population composition rather than the intrinsic evolution of individual galaxies \citep{Lilly2016, Chenyy2024}. Moreover, the existence of a minority population of extended quenched galaxies further suggests that multiple quenching pathways may operate. Dynamical heating associated with processes such as violent accretion and mergers can modify galaxy structures after quenching \citep{Chenyy2024, Maql2026}, further complicating the interpretation of observations. Thus, whether compact star-forming galaxies represent the dominant progenitors of compact quenched galaxies still remains an open question.

To quantify the connection between structural properties and quenching, previous studies have adopted various measures of galaxy structure, including effective stellar mass surface density ($\Sigma_{\rm e}$; e.g., \citealp{Cheung2012, Barro2017ApJ, Mosleh2017, Whitaker2017, Yang2025}) and $\Sigma_1$ \citep{Cheung2012, fang2013, Barro2017ApJ, Mosleh2017, Tacchella2017, Wang2018, Guo2021}. While global density measures capture overall compactness, $\Sigma_1$ directly quantifies the central stellar mass concentration, which is closely linked to bulge growth and central velocity dispersion \citep{fang2013, Woo2015, Barro2017ApJ}. Compared with global density measures such as $\Sigma_{\rm e}$, $\Sigma_1$ shows a tighter connection with star formation activity \citep{Cheung2012, Mosleh2017}. Therefore, we adopt $\Sigma_1$ as a physically motivated tracer of galaxy compactness. Following recent machine-learning approaches that quantify the relative predictive power of different galaxy properties \citep{Bluck2020b, Bluck2020, Chenyy2020,Chenyy2021a,Chenyy2021b,Bluck2022}, we employ a Random Forest framework to assess the relative importance of $\Sigma_1$ and other galaxy properties in predicting quenching.

Galaxies form and evolve within dark matter halos, and environmental effects may provide additional influences on their evolutionary pathways \citep{WhiteRees1978,Mo2010}. The structure of galaxies may partially reflect the properties of their host halos, as dark matter halos provide the gravitational potential that drives baryons to accumulate and reach the conditions required for star formation \citep{Chenyy2024}, and both dark matter halos and baryonic gas acquire angular momentum through tidal torques and merger processes. Furthermore, halo assembly history may imprint correlations between galaxy properties and large-scale structure through assembly bias \citep{Gao2005, Zentner2014, Chenyy2020, Lyu2023, Wangk2023a, Wangk2023b, Lyu2024, Wangk2025, Zhang2025}. Although massive galaxies are generally considered to be primarily regulated by internal processes, their structural diversity may also reflect differences in their halo environments and assembly histories (see e.g. \citealp{Mo2024}). Therefore, simultaneously characterizing galaxy structure and halo environment is essential for obtaining a complete picture of massive galaxy evolution.

Recent advances with the James Webb Space Telescope (JWST) provide unprecedented opportunities to study the interplay between galaxy structure, star formation, and environment at cosmic noon. JWST near-infrared imaging enables detailed measurements of rest-frame optical structures at $1\leq z<3$. The COSMOS-Web \citep{Casey2023} is the largest contiguous JWST NIRCam survey field to date, combines deep imaging with extensive ancillary data, enabling statistical studies of galaxy evolution and large-scale environments with reduced cosmic variance.

The main goal of this paper is to investigate the connection between galaxy structural properties, quenching, and large-scale environments at cosmic noon using JWST/COSMOS-Web observations. In Section~\ref{sec_data}, we describe the data and galaxy sample selection used in this work. In Section~\ref{sec_method}, we present the methodology, including measurements of galaxy structural properties, structural classification, clustering analysis, and the Random Forest approach. In Section~\ref{sec_results}, we present our main results on the relative importance of galaxy properties in predicting quenching, the properties of compact and extended galaxy populations, and differences in their clustering properties. In Section~\ref{sec_discussion}, we discuss the implications of our results. Finally, we summarize our conclusions in Section~\ref{sec_summary}. Throughout this paper, we assume a flat $\Lambda$CDM cosmology with $\rm \Omega_m$ = 0.307,  $\rm \Omega_b$ = 0.048, $\rm \Omega_\Lambda$ = 0.693 and $h = 0.678$.  Here, $h = \rm H_0/(100$ km $\rm s^{-1}$ $\rm Mpc^{-1})$ and $\rm H_0$ is the Hubble constant \citep{Planck2016}.

\section{Observational data}\label{sec_data}
\subsection{The COSMOS-Web catalog}\label{sec_cosmo_cat} 
The COSMOS-Web survey \citep{Casey2023} is a JWST Cycle 1 program covering 0.54 deg$^2$ in the COSMOS field with four NIRCam filters (F115W, F150W, F277W, F444W), reaching a 5$\sigma$ depth of $\sim$27.5–28.2 mag, and supplemented by MIRI F770W imaging over a non-contiguous $\sim$0.19 deg$^2$ area with depths of $\sim$25–26 mag. The observations were obtained across multiple epochs between 2023 and 2024, and the data reduction follows standard JWST pipelines with detailed processing described in \cite{Franco2026} and \cite{Harish2025}. 

The survey is combined with extensive ancillary multi-wavelength data in COSMOS, enabling the construction of the COSMOS-Web photometric catalog\footnote{https://cosmos2025.iap.fr/catalog.html} \citep{Shuntov2025}, which contains $\sim780,000$ sources. Photometry and structural parameters are derived using SourceXtractor++ with PSF-convolved Sérsic profile fitting, which provides robust structural parameters, including the Sérsic effective radius ($R_{\rm e}$), defined as the radius enclosing half of the galaxy's total light, and the Sérsic index ($n$) used in this work. By taking advantage of JWST’s high spatial resolution, this approach enables reliable source deblending and improved photometric measurements, making the dataset a powerful resource for studying galaxy evolution across cosmic time.

Photometric redshifts are derived using $\bf\rm LePHARE$ \citep{Arnouts2002, Ilbert2006} through spectral energy distribution (SED) fitting based on \cite{Bruzual2003} stellar population models and a range of star formation histories \citep{Ilbert2015}. The fitting includes dust attenuation (with multiple extinction laws and varying E(B$-$V)), emission lines, and intergalactic medium absorption. Redshifts ($z$) are taken as the median of the posterior probability distributions, with uncertainties from the 16th–84th percentiles. The photo-$z$ accuracy is evaluated against $\sim$12,000 spectroscopic redshifts in COSMOS, showing good agreement.

We adopt the non-parametric star formation histories (SFHs) and the corresponding star-formation rate (SFR), stellar mass and stellar ages derived from SED fitting with $\bf\rm CIGALE$ \citep{Boquien2019}, provided in Column 4 of the catalog (“CIGALE Physical Parameters and Non-Parametric SFH”). Here, the SFR is calculated over the look-back time of the last 100 Myr, the stellar age is defined as the look-back time between the epoch of observation and the epoch at which a galaxy formed 50\% of its total stellar mass. Using the observed redshift $z_{\rm obs}$ together with the stellar age, we further calculate the formation redshift ($z_{\rm form}$) for each galaxy by converting the stellar age into the corresponding cosmic epoch:
\begin{equation}
t_{\rm age}=t(z_{\rm obs})-t(z_{\rm form}),
\end{equation}
where $t(z)$ is the cosmic age at redshift $z$. Equivalently, $z_{\rm form}$ corresponds to the redshift at which the galaxy had formed 50\% of its total stellar mass.

\subsection{Galaxy sample selection}\label{sec_samps}
Following \cite{Paquereau2025}, to ensure a representative galaxy sample for clustering measurements, we applied a series of cleaning steps. Stars were removed based on their better fits to stellar or brown dwarf templates in $\bf\rm LePHARE$, combined with a compactness criterion. We also excluded sources with sizes smaller than twice the FWHM in the $F_{\rm 115W}$ band, as well as objects flagged hot pixels or other image artifacts. Regions around bright stars in both NIRCam and optical (HSC) images were masked, removing a small fraction of the sample ($\sim$17\% in total).

We then imposed a magnitude limit of $m_{\rm 444W}$=27.75, following \cite{Paquereau2025}, defined from completeness tests against the deeper PRIMER-COSMOS survey and adopting an 80\% completeness threshold. Stellar mass completeness was estimated following \cite{Pozzetti2010}, by computing, for each galaxy, the limiting mass it would have at the survey magnitude limit in the $F_{\rm 444W}$ band. The final stellar mass completeness in each redshift bin is defined as the mass above which 90\% of galaxies are expected to be complete, increasing from $\log(M_*/M_\odot)\sim7.5$ at $z\sim1$ to $\log(M_*/M_\odot)\sim8$ at $z\sim3$. We define the parent sample as galaxies with stellar masses above the corresponding completeness limits. Over the cosmic noon period, we adopt a conservative mass cut of $\log(M_*/M_\odot)>9$.

We further divide the parent sample into two redshift bins, \(1.0\leq z<1.7\) and \(1.7\leq z<3.0\), to examine whether the observed galaxy properties and their halo connections show systematic variations across different epochs of cosmic noon. The two redshift intervals correspond to time spans of approximately 2 and 1.6 Gyr, respectively. We excluded galaxies for which less than 30\% of the photometric-redshift probability distribution lies within the redshift bin of interest, i.e., \(P(z_{\min}\leq z\leq z_{\max})<0.3\), where \(z_{\min}\) and \(z_{\max}\) are the lower and upper boundaries of the redshift bin, respectively.

In this work, we investigate the evolution of massive galaxies with $\log(M_*/M_\odot)>10.5$. These galaxies are selected from the parent sample described above and are referred to as the full samples throughout this work. They are predominantly central galaxies in their host dark matter halos, with satellite fractions estimated from the halo occupation distribution (HOD) method to be $\lesssim 7\%$ \citep{Paquereau2025}.

To separate star-forming and quenched galaxies from the full samples, we adopt the NUV–r–J color-color diagram, which has been widely used in studies of galaxies at intermediate to high redshifts \citep[e.g.,][]{Ilbert2013, Yang2025}:
\begin{equation}
    \begin{aligned}
        M_{\rm NUV} - M_{\rm r} &> 3 \  (M_{\rm r} - M_{\rm J}), \\
        M_{\rm NUV} - M_{\rm r} &> 3.1,
    \end{aligned}
    \label{eq_sfq}
\end{equation}
where $M_{\rm NUV}$, $M_{\rm r}$ and $M_{\rm J}$ are the absolute magnitudes in the NUV, r, and J bands, respectively, estimated from SED fitting. 

\begin{figure}
    \centering        
    \includegraphics[width=1\linewidth]{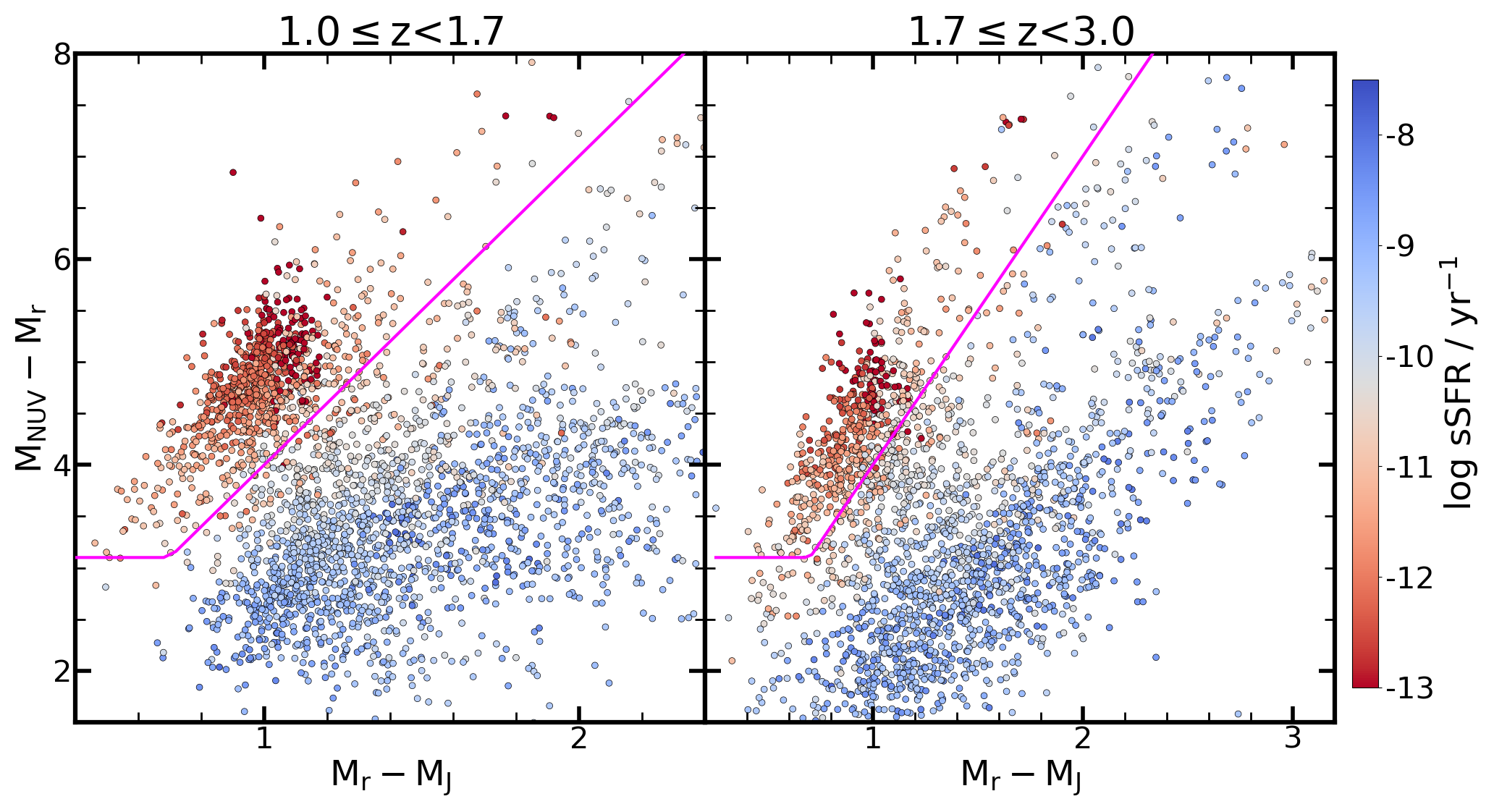}
    \caption{Rest-frame \(\rm NUV-r-J\) color–color diagram of galaxies at \(1.0\leq z<1.7\) (left) and \(1.7\leq z<3.0\) (right). Symbols are colored by specific star formation rate (sSFR). The magenta line denotes the \(\rm NUV-r-J\) selection boundary (Equation \ref{eq_sfq}), partitioning quenched galaxies (upper left) from star-forming galaxies (lower right). Galaxies are further classified via an additional sSFR cut: quenched galaxies satisfy \(\log\ {\rm sSFR} < \log\ 0.2/t(z_{\rm obs})\), and star-forming galaxies satisfy \(\log\ {\rm sSFR} > \log\ 0.2/t(z_{\rm obs})\), with \(t(z_{\rm obs})\) corresponding to the age of the Universe at the corresponding redshift.}
    \label{fig_sfq}
\end{figure}

The specific star-formation rate (sSFR) was also commonly applied to separate star-forming and quenched galaxies, with quenched galaxies defined as having ${\rm log}\ {\rm sSFR} < {\rm log}\ 0.2/t(z_{\rm obs})$, where $t(z_{\rm obs})$ is the age of the Universe at the corresponding redshift \citep[e.g.,][]{Pacifici2016, Carnall2023, Yang2025}. We further apply an sSFR criterion together with the color--color selection to construct conservative star-forming and quenched samples. Specifically, the color-selected star-forming galaxies are required to have $\rm log \ sSFR>{\rm log}\ 0.2/t(z_{\rm obs})$, while the color-selected quenched galaxies are required to have $\rm log \ sSFR<{\rm log}\ 0.2/t(z_{\rm obs})$. The separation between star-forming and quenched galaxies in our sample is shown in Figure~\ref{fig_sfq}. This combined selection leaves a small fraction of galaxies with inconsistent classifications, including blue galaxies with low sSFRs and red galaxies with high sSFRs. We exclude these ambiguous populations and define the resulting star-forming and quenched samples as our fiducial samples. The impact of these ambiguous populations is assessed by including them in the galaxy samples and adopting alternative star-forming/quenched classification criteria. The corresponding tests are presented in Appendix~\ref{sec_alter}.

To ensure reliable measurements of the central stellar density, we only select galaxies with \(0.5<n<6\) and \(R_{\rm e}>0.3\,\rm kpc\) (see Section \ref{sec_sigma1}). We further divide the fiducial star-forming and quenched samples into compact and extended populations based on their structural properties (see Section \ref{sec_gal_pro}). The detailed properties of each galaxy subsample selected using the fiducial criteria, including the sample size, galaxy bias, and inferred halo mass in each redshift bin, are summarized in Table~\ref{tab_samps}.

\begin{table*}[t]
\centering
\small
\caption{Number of galaxies, galaxy biases, and inferred halo masses of different galaxy subsamples selected using the fiducial criteria.}
\label{tab_samps}

\begin{threeparttable}
\begin{tabular*}{\textwidth}{@{\extracolsep{\fill}}c|ccc|ccc}

\hline
\hline

Subsample\tnote{(a)}
& \multicolumn{3}{c|}{$1\leq z<1.7$}
& \multicolumn{3}{c}{$1.7\leq z<3$}
\\

& $N_{\rm gal}$\tnote{(b)}
& $b_{\rm g}$\tnote{(c)}
& ${\rm log} M_{\rm h}/{\rm M_{\odot}}$\tnote{(d)}
& $N_{\rm gal}$
& $b_{\rm g}$
& ${\rm log} M_{\rm h}/{\rm M_{\odot}}$
\\

\hline

Full
& 2,841
& 1.81$^{+0.25}_{-0.29}$ 
& 12.4$^{+0.23}_{-0.35}$
& 2,304
& 3.11$^{+0.54}_{-0.66}$ 
& 12.48$^{+0.24}_{-0.39}$
\\

Star-forming (SF)
& 1,737
& 1.2$^{+0.25}_{-0.32}$ 
& 11.47$^{+0.48}_{-1.12}$
& 1,435
& 2.84$^{+0.47}_{-0.57}$ 
& 12.33$^{+0.24}_{-0.38}$
\\

Quenched (Q)
& 858
& 2.12$^{+0.5}_{-0.66}$ 
& 12.68$^{+0.33}_{-0.72}$
& 512
& 5.06$^{+1.46}_{-2.11}$ 
& 13.15$^{+0.3}_{-0.75}$
\\

Extended star-forming (eSF)
& 1,274
& 1.02$^{+0.21}_{-0.27}$ 
& 10.81$^{+0.62}_{-1.64}$
& 989
& 2.2$^{+0.67}_{-0.99}$ 
& 11.83$^{+0.46}_{-1.49}$
\\

Compact star-forming (cSF)
& 463
& 0.36$^{+0.59}_{-0.16}$ 
& 9.17$^{+1.4}_{-0.0}$
& 446
& 2.0$^{+0.76}_{-1.38}$ 
& 11.63$^{+0.59}_{-2.47}$
\\

Extended quenched (eQ)
& 49
& --
& --
& 31
& --
& --
\\

Compact quenched (cQ)
& 809
& 1.93$^{+0.47}_{-0.62}$ 
& 12.44$^{+0.36}_{-0.84}$
& 481
& 4.78$^{+1.91}_{-3.79}$
& 13.04$^{+0.4}_{-3.47}$
\\

\hline
\hline

\end{tabular*}

\begin{tablenotes}[para,flushleft]
\footnotesize
\setlength{\hsize}{\textwidth}

\item[(a)]
Massive galaxy subsamples used in this work.

\item[(b)] 
The galaxy number of each subsample. 

\item[(c)]
The inferred galaxy bias of each subsample.

\item[(d)] 
The inferred $M_{\rm 200m}$ halo mass of each subsample.

\end{tablenotes}
\end{threeparttable}

\end{table*}

\subsection{Environmental density}
We adopt the environmental density map reconstructed by \cite{Hatamnia2025}, who applied the weighted kernel density estimation (wKDE) method to reconstruct the galaxy number density field using a COSMOS-Web galaxy sample with a stellar mass limit of $\log(M_\star/M_\odot)\geq 8$. The dimensionless galaxy overdensity is defined as
\begin{equation}
    \delta = \frac{\sigma^{\rm s}(\mathbf{X})-\bar{\sigma}^{\rm s}}{\bar{\sigma}^{\rm s}},
\end{equation}
where $\mathbf{X}$ denotes the angular coordinates of the target galaxy, $\sigma^{\rm s}(\mathbf{X})$ represents the weighted density at the galaxy position, and $\bar{\sigma}^{\rm s}$ is the mean weighted density within the corresponding redshift slice. The density field is reconstructed in COSMOS-Web photometric redshift slices by accounting for galaxy redshift uncertainties and correcting for survey geometry and masking effects \citep{Hatamnia2025}. Previous comparisons based on simulations \citep{Darvish2015} have demonstrated that both wKDE and Voronoi tessellation can effectively recover the underlying density field, with wKDE exhibiting greater robustness against shot noise and sparse sampling.

\section{Methods of analysis}\label{sec_method}
\subsection{Stellar mass and redshift matching}\label{sec_matching}
To minimize the impact of different stellar mass and redshift distributions among galaxy populations, we construct stellar mass- and redshift-matched samples using an iterative reweighting procedure.

For the target galaxy sample, we assign an individual weight to each galaxy and iteratively adjust these weights to match the stellar mass and redshift distributions of the base sample \citep{Zhang2025}. Starting from equal weights, the weights are updated in each iteration according to
\begin{equation}
w^{(i+1)}
=
w^{(i)}
\frac{{\rm PDF}_{\rm base}(M_\star)}
{{\rm PDF}_{w}^{(i)}(M_\star)}
\frac{{\rm PDF}_{\rm base}(z)}
{{\rm PDF}_{w}^{(i)}(z)},
\end{equation}
where \(w^{(i)}\) represents the weight of each galaxy after the \(i\)-th iteration. The terms \({\rm PDF}_{\rm base}(M_\star)\) and \({\rm PDF}_{\rm base}(z)\) denote the stellar mass and redshift distributions of the base sample, respectively, while \({\rm PDF}_{w}^{(i)}(M_\star)\) and \({\rm PDF}_{w}^{(i)}(z)\) represent the corresponding weighted distributions of the target sample at the \(i\)-th iteration. The PDF ratios are evaluated in the corresponding stellar mass and redshift bins, and each target galaxy's weight is updated according to the ratios in the bins containing that galaxy.

The stellar mass and redshift reweighting steps are alternately repeated until the weighted target sample converges to the base stellar mass and redshift distributions. In our analysis, we perform 10 iterations for the target galaxy sample, which is sufficient to achieve convergence. The final weights will be used in measuring the clustering signals (see Section~\ref{sec_clustering}).

When comparing the clustering properties of the full, star-forming, and quenched samples, we adopt the full sample within each redshift bin as the base sample for the matching procedure. For comparisons among extended star-forming, compact star-forming, and compact quenched galaxies, we use the compact star-forming sample in each redshift bin as the base sample. For illustration, the weighted and unweighted distributions of the full, star-forming, and quenched samples are shown in Figure \ref{fig_dis1} in the Appendix~\ref{sec_weighting}.

\subsection{Galaxy clustering measurements}\label{sec_clustering}
We measure the galaxy clustering using the two-point angular auto-correlation function (ACF) and adopt the estimator of Landy \& Szalay \citep{Landy1993}, defined as follows:
\begin{equation}\label{eq_ls}
    \omega_{\rm ACF}(\theta) = \frac{GG(\theta) - 2\ GR(\theta) + RR(\theta)}{RR(\theta)},
\end{equation}
where $GG$, $GR$, and $RR$ represent the normalized numbers of galaxy-galaxy, galaxy-random, and random-random pairs, respectively. For the galaxy-galaxy and galaxy-random pair counts ($GG$ and $GR$), we incorporate the weights of the galaxy samples when computing the pair counts. The random catalog is assigned unit weight. In this work, we use the Corrfunc package\footnote{https://github.com/manodeep/corrfunc} \citep{Sinha2017} to efficiently compute the pair counts.

We construct a random catalog that matches the survey geometry and angular selection function of the parent sample (Section~\ref{sec_samps}). To reduce Poisson noise, the random catalog contains objects that is 15 times of the parent sample. Random sources are distributed uniformly across the effective survey area of the parent sample, while applying the same survey masks as those used for the data. This procedure ensures that the random catalog accurately reproduces the observational footprint and selection effects of the survey.

We estimate the values and uncertainties of the ACF using the jackknife resampling method. The observed field is divided into \(6\times5\) spatial subregions. For each jackknife realization, one subregion is removed and the ACF is recalculated, yielding \(\omega_{{\rm obs},k}(\theta)\), where \(k\) denotes the \(k\)-th jackknife realization. The final observed ACF signal in each angular bin is obtained by averaging over all jackknife realizations,
\begin{equation}
    \omega_{\rm obs}(\theta)=\frac{1}{N}\sum_{k=1}^{N}\omega_{{\rm obs},k}(\theta),
\end{equation}
where \(N\) is the number of jackknife realizations. The uncertainty of $\omega_{\rm obs}(\theta)$ is estimated from the jackknife variance:
\begin{equation}
    {\rm Var}(\theta) =
    \frac{N-1}{N}
    \sum_{k=1}^{N}
    \left(\omega_{{\rm obs},k}(\theta)-\omega_{\rm obs}(\theta)\right)^2,
\end{equation}
The corresponding uncertainty is given by \(\sigma(\theta)=\sqrt{{\rm Var}(\theta)}\).

The COSMOS-Web survey covers an area of only about 0.54 $\rm deg^2$, which can introduce the integral constraint bias and lead to an underestimation of the observed correlation function. To correct for this effect, following \cite{Ito2021}, the integral constraint $C$ is estimated as
\begin{equation}
    C=\frac{\Sigma_i \theta_i^{1-\gamma} RR(\theta_i)}{\Sigma_i RR(\theta_i)}.
\end{equation}
where $\theta_i$ denotes the \(i\)-th angular separation bin, and $\omega_{\rm obs}(\theta)$ is corrected according to
\begin{equation}\label{eq_cacf}
    \omega_{\rm ACF}(\theta) = \omega_{\rm obs}(\theta)\frac{\theta^{1-\gamma}}{\theta^{1-\gamma}-C}.
\end{equation}

In our analysis, the multiplicative integral constraint correction factor becomes non-physical in the largest angular bin, we therefore do not apply the multiplicative correction to this bin and use the measured correlation amplitude directly.

\subsection{Galaxy bias and halo mass}\label{sec_bias_Mh}

The ACF is commonly described by a power-law form,
\begin{equation}\label{eq_pl}
\omega_{\rm ACF}(\theta) = A_{\omega}\theta^{1-\gamma},
\end{equation}
where $A_{\omega}$ is the clustering amplitude and $\gamma$ is the slope of the correlation function. 

To determine $\gamma$, we first measure the observed ACF of the parent sample, which provides the highest signal-to-noise ratio compared with individual subsamples. The parent sample is assigned unit weight. Since the integral constraint correction depends on the assumed $\gamma$, we adopt an iterative procedure to obtain a self-consistent value of $\gamma$. Starting from an initial value of $\gamma=1.8$ \citep{Coupon2012, Ito2021}, we calculate the integral constraint correction and fit the corrected correlation function with a power-law (Equation~\ref{eq_pl}), allowing both $A_{\omega}$ and $\gamma$ to vary. The updated value of $\gamma$ is then used for the next iteration. This procedure is repeated for 10 times, after which $\gamma$ converges to a stable value. The final $\gamma$ is adopted for the clustering analysis of the other galaxy subsamples at the same redshift bin (see Table~\ref{tab_acf_fit}). The fitting to the power-law is performed over angular scales of \(\theta>0.01^\circ\) for the corrected ACF.

The three-dimensional spatial correlation function is also assumed to follow a power-law form,
\begin{equation}
\xi(r)=\left(\frac{r}{r_0}\right)^{-\gamma},
\end{equation}
where $r_0$ is the correlation length. We then employ the Limber inversion \citep{Peebles1980} to convert the projected two-dimensional ACF into the spatial correlation function:
\begin{equation}
A_{\omega}
=
C_{\gamma}r_0^{\gamma}
\frac{
\int_0^{\infty}
F(z)
D_{\theta}(z)^{1-\gamma}
N(z)^2
g(z)
(1+z)^{-\gamma}
dz
}
{
\left[\int_0^{\infty}N(z)dz\right]^2
},
\end{equation}
where
\begin{equation}
g(z)
=
\frac{H_0}{c}
(1+z)^2
\{
1+\Omega_m z
+\Omega_{\Lambda}\left[(1+z)^{-2}-1\right]
\}^{1/2},
\end{equation}
and
\begin{equation}
C_{\gamma}
=
\frac{
\sqrt{\pi}\
\Gamma[(\gamma-1)/2]
}
{
\Gamma(\gamma/2)
}.
\end{equation}

Here, $c$ is the light speed, $D_{\theta}(z)$ is the angular diameter distance, $N(z)$ is the redshift distribution of the target galaxy sample, and $\Gamma$ denotes the Gamma function. The function
\begin{equation}
F(z)=\left(\frac{1+z}{1+\bar{z}}\right)^{-(3+\epsilon)}
\end{equation}
describes the redshift evolution of the spatial correlation function, where we adopt $\epsilon=-1.2$ following \cite{Roche1999}, and $\bar{z}$ is the mean redshift of the target galaxy sample.

Assuming a linear galaxy bias model, the galaxy bias of the target sample is estimated by comparing its spatial correlation function with the dark matter correlation function on scales dominated by the two-halo term. Specifically, we compute
\begin{equation}
b_{\rm target}\label{eq_gb}
=
\left\langle
\sqrt{
\frac{
\xi(r)
}
{
\xi_{\rm DM}(r,\bar{z})
}
}
\right\rangle_{3<r<5~{\rm cMpc}/h},
\end{equation}
where the bias is averaged over the comoving scale range of $3-5~{\rm cMpc}/h$. The dark matter correlation function, $\xi_{\rm DM}(r,\bar{z})$, is computed at the mean redshift $\bar{z}$ using the \texttt{COLOSSUS} package\footnote{https://pypi.org/project/colossus/} \citep{Diemer2018ApJS}.

The halo mass of the target galaxy sample is inferred by interpolating the galaxy bias derived from Equation~(\ref{eq_gb}) onto the halo mass--bias relation of \cite{Tinker2010}, computed using the \texttt{COLOSSUS} package. Here, the halo mass is defined as $M_{200\mathrm{m}}$, i.e., the mass enclosed within a radius $R_{200\mathrm{m}}$ where the mean mass density is 200 times the mean matter density of the Universe.

\subsection{Random Forest framework for quantifying the predictive power of galaxy properties}\label{sec_rf}

To investigate the galaxy properties associated with quenching, we employ a Random Forest (RF) classifier to distinguish star-forming and quenched galaxies. The input feature set consists of five galaxy properties: $\Sigma_1$, $z_{\rm form}$, $M_*$, $1+\delta$, and $R_{\rm e}$. Galaxies with unavailable measurements are excluded from the analysis. For each redshift bin, the galaxy sample is randomly divided into training and test sets with a ratio of 4:1. The RF classifier consists of 300 decision trees, with a maximum tree depth of 6 and a minimum of 20 galaxies in each terminal leaf. Class-balanced weights are adopted to compensate for the imbalance between star-forming and quenched galaxies. 


To quantify the relative importance of different galaxy properties, we train an RF classifier for each redshift bin using all five input features and interpret the trained model using SHAP (SHapley Additive exPlanations) scores computed with the TreeExplainer algorithm. The SHAP score of each feature quantifies its contribution to the predicted probability for an individual galaxy. Because several galaxy properties are mutually correlated (e.g., $M_*$, $\Sigma_1$, and $R_{\rm e}$), we evaluate the importance of both individual parameters and their pairwise combinations. For each feature group, the SHAP scores of the constituent features are summed for every galaxy, and the group importance is defined as the mean absolute SHAP score over the test sample. The feature groups are then ranked according to this quantity, allowing a direct comparison of the relative contributions of different galaxy properties to quenching. The Random Forest classification is performed using the \texttt{scikit-learn}\footnote{https://pypi.org/project/scikit-learn/} and \texttt{SHAP}\footnote{https://pypi.org/project/shap/} Python packages.


\subsection{Measurement of central stellar mass density}\label{sec_sigma1}
We compute $\Sigma_1$ by assuming that the stellar mass follows a Sérsic profile. The enclosed stellar mass fraction within a projected radius $r$ is calculated using the incomplete gamma function,
\begin{equation}
f(<r)=
\frac{\gamma\left(2n,b_{\rm n}(r/R_{\rm e})^{1/n}\right)}
{\Gamma(2n)},
\end{equation}
where $R_{\rm e}$ is the effective radius, $n$ is the Sérsic index, and $b_{\rm n}$ is determined by the condition that $R_{\rm e}$ encloses half of the total stellar mass. We adopt the approximation \citep{Prugniel1997, Terzi2005}
\begin{equation}
b_{\rm n}=2n-\frac{1}{3}+\frac{0.009876}{n}.
\end{equation}

The stellar mass enclosed within 1 kpc is then calculated as
\begin{equation}
M(<1\,\mathrm{kpc}) = f(<1\,\mathrm{kpc}) \, M_\star,
\end{equation}
and the central stellar mass density is defined as
\begin{equation}
\Sigma_1 = \frac{M(<1\,\mathrm{kpc})}{\pi (1\,\mathrm{kpc})^2}.
\end{equation}

We restrict our analysis to galaxies with $0.5<n<6$ and $R_{\rm e}>0.3\,\mathrm{kpc}$ to minimize uncertainties in the inferred $\Sigma_1$. Extreme Sérsic indices can lead to unstable estimates of the enclosed mass fraction, while galaxies with very small effective radii are more susceptible to resolution effects and uncertainties in structural parameter measurements.

\subsection{Classification of compact and extended galaxies}\label{sec_fit_ridge}
We define compact and extended populations based on the structural sequence of quenched galaxies. Following previous studies \citep{Barro2017ApJ, Wang2018a}, we characterize the structural sequence of quenched galaxies by fitting a ridgeline in the $M_*-\Sigma_1$ plane within each redshift bin:
\begin{equation}
\log \Sigma_1 = a + b(\log M_* - 11),
\end{equation}
where $a$ and $b$ are determined through a least-squares fit. Unlike a fixed $\Sigma_1$ threshold, the ridgeline method accounts for the mass dependence of galaxy structure and allows us to quantify compactness relative to the quenched population at a given stellar mass. We define the offset from this relation as
\begin{equation}
\Delta \Sigma_1 = \log \Sigma_1 - \log \Sigma_{1,\mathrm{ridge}}(M_*),
\end{equation}
which quantifies the relative compactness of each galaxy at fixed stellar mass and redshift bin.

The scatter of the quenched structural sequence is quantified using the median absolute deviation (mad) of $\Delta\Sigma_1$:
\begin{equation}
\sigma_{\rm ridge}=1.4826\ \sigma_{\rm mad}(\Delta\Sigma_1),
\end{equation}
where the mad provides a robust estimate of the scatter around the quenched ridgeline, and the factor 1.4826 scales it to the equivalent standard deviation for a Gaussian distribution. We then classify star-forming and quenched galaxies as compact or extended according to their offsets from the quenched ridgeline:
\begin{equation}
\begin{aligned}
{\rm Compact}:&\quad \Delta\Sigma_1 \geq -2\sigma_{\rm ridge},\\
{\rm Extended}:&\quad \Delta\Sigma_1 < -2\sigma_{\rm ridge}.
\end{aligned}
\end{equation}

\section{Results}\label{sec_results}
\subsection{Relative importance of galaxy properties in predicting galaxy quenching}\label{sec_qf}

\begin{figure*}
    \centering
    \includegraphics[width=1\linewidth]{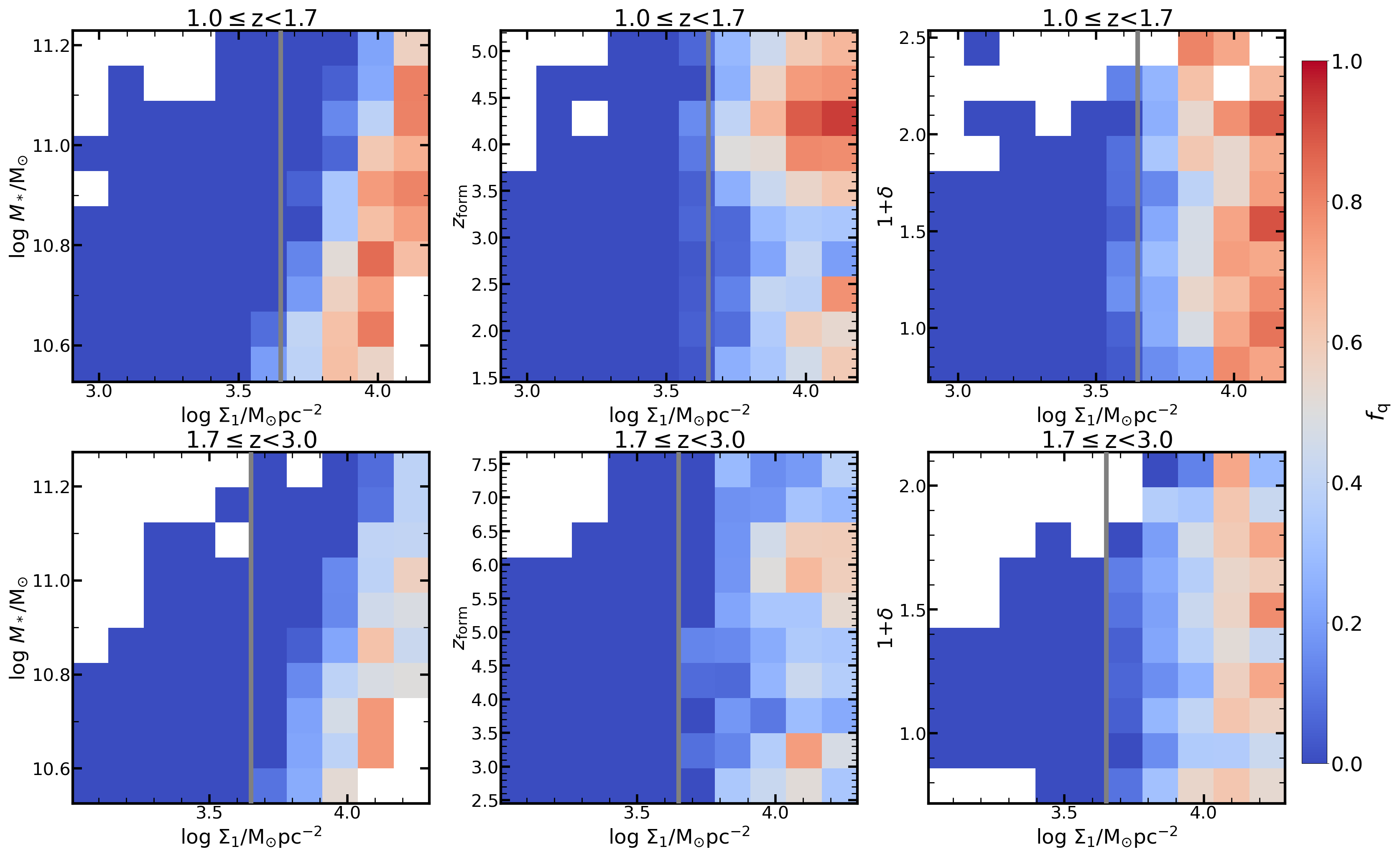}
    \caption{Two-dimensional distributions of the quenched fraction in different parameter spaces. The upper and lower rows correspond to the redshift bins \(1\leq z<1.7\) and \(1.7\leq z<3\), respectively. The left, middle, and right columns show the \(\Sigma_1-M_*\), \(\Sigma_1-z_{\rm form}\), and \(\Sigma_1-(1+\delta)\) parameter spaces, respectively. The vertical grey line in each panel corresponds to  \({\rm log\ }\Sigma_1/{\rm M_{\odot}pc^{-2}}=3.65\). The color scale indicates the quenched fraction in each parameter bin.}
    \label{fig_fq_2d}
\end{figure*}

The quenching of massive galaxies is closely associated with both their internal properties and surrounding environments. Previous studies have suggested that stellar mass, central stellar density, galaxy age, and local environment may all correlate with the transition from star-forming to quenched populations \citep{Peng2010, fang2013, Woo2013a}. We therefore first investigate the two-dimensional distributions of the quenched fraction ($f_{\rm q}$) as a function of $M_\star$, $\Sigma_1$, $z_{\rm form}$, and $(1+\delta)$ in both redshift bins. The quenched fraction is defined as $f_{\rm q}=N_{\rm Q}/N_{\rm all}$, where $N_{\rm Q}$ and $N_{\rm all}$ denote the numbers of quenched galaxies and all galaxies within a given bin, respectively. The results are shown in Figure~\ref{fig_fq_2d}.

The quenched fraction exhibits a strong dependence on $\Sigma_1$ in both redshift bins.  In contrast, $f_{\rm q}$ shows only a weak dependence on $M_\star$ within our massive galaxy sample. Galaxies with low $\Sigma_1$ predominantly have very low $f_{\rm q}$, while regions with high $\Sigma_1$ span a broader range of quenched fractions, including parameter bins with substantially elevated $f_{\rm q}$. A threshold-like behavior is observed around ${\rm log\ }\Sigma_1/{\rm M_{\odot}pc^{-2}}\simeq3.6-3.7$, marked by the grey vertical line in Figure \ref{fig_fq_2d}: above this characteristic \(\Sigma_1\), bins with high \(f_{\rm q}\) become progressively more prevalent. This behavior is broadly consistent with previous findings that quenched galaxies preferentially occupy regions of high $\Sigma_1$ \citep{fang2013, Woo2015, Barro2017ApJ}. \cite{fang2013} showed that galaxies with ${\rm log\ }\Sigma_1/{\rm M_{\odot}pc^{-2}}\gtrsim3.5$ host the majority of red galaxies, while \cite{Barro2017ApJ} found that a mass-dependent compactness criterion, ${\rm log\ }\Sigma_1/{\rm M_{\odot}pc^{-2}}-0.65(\log M_\ast/{\rm M_{\odot}}-10.5)\gtrsim3.5$, efficiently selects compact star-forming and quenched galaxies at $z\gtrsim0.5$.

The dependence of $f_{\rm q}$ on $z_{\rm form}$ is mainly observed among galaxies with high $\Sigma_1$ and varies between the two redshift bins. In the lower-redshift bin ($1\leq z<1.7$), $f_{\rm q}$ increases toward higher $z_{\rm form}$ for galaxies with high $\Sigma_1$, with the highest quenched fractions found in bins combining high $\Sigma_1$ and high $z_{\rm form}$. In the higher-redshift bin ($1.7\leq z<3$), this trend becomes weaker, although a mild preference for higher $f_{\rm q}$ at larger $z_{\rm form}$ remains visible among high-$\Sigma_1$ galaxies. In both redshift bins, $f_{\rm q}$ does not exhibit a clear dependence on $(1+\delta)$.

\begin{figure*}
    \centering
    \includegraphics[width=1\linewidth]{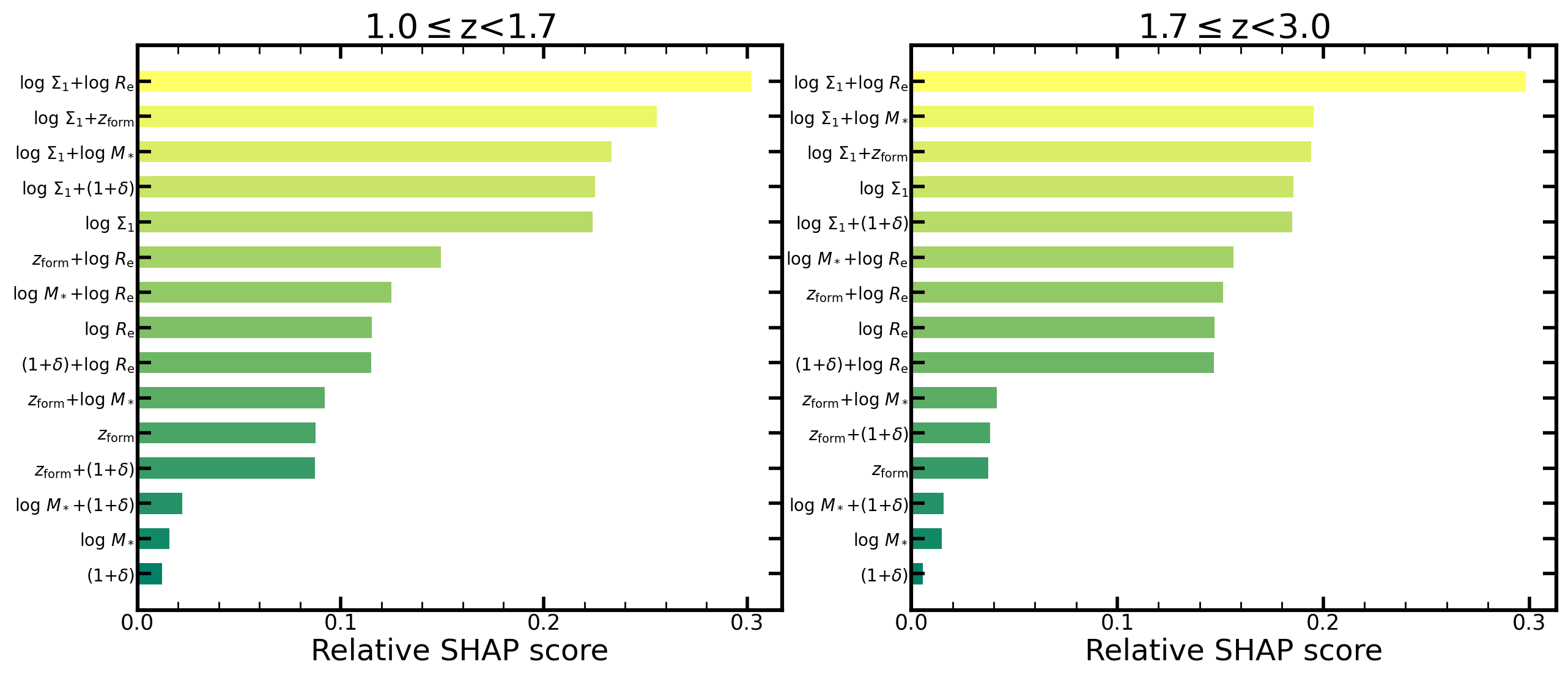}
    \caption{Relative importance of different feature sets for predicting galaxy quenching based on SHAP scores. The SHAP values are derived from Random Forest models trained with all five features ($\Sigma_1$, $z_{\rm form}$, $M_\ast$, $R_{\rm e}$, and $(1+\delta)$). The left and right panels correspond to \(1\leq z<1.7\) and \(1.7\leq z<3\), respectively. The x-axis shows the relative SHAP score, and the y-axis lists different feature sets. The horizontal bars represent the SHAP importance of each combination.}
    \label{fig_SHAP}
\end{figure*}

Applying the Random Forest analysis described in Section~\ref{sec_rf}, we evaluate the relative importance of different galaxy properties in predicting galaxy quenching using SHAP scores. Figure~\ref{fig_SHAP} presents the SHAP importance of individual parameters and feature combinations in the two redshift bins. In both redshift bins, among the five input features considered here, $\Sigma_1$ emerges as the most important predictor in distinguishing quenched and star-forming galaxies. Feature combinations containing $\Sigma_1$ consistently exhibit higher SHAP scores than those without $\Sigma_1$. $R_{\rm e}$ is the second most important parameter, likely reflecting its connection with $\Sigma_1$, followed by $z_{\rm form}$, $M_*$, and $(1+\delta)$.


These results are consistent with previous studies demonstrating the close connection between $\Sigma_1$ and quenching across cosmic time \citep{fang2013, Barro2017ApJ, Whitaker2017}. Other galaxy properties provide additional but relatively limited predictive information compared with $\Sigma_1$. This motivates us to investigate whether galaxies with different structural states represent distinct stages in galaxy evolution.

\subsection{Structural diversity of star-forming and quenched galaxies}\label{sec_gal_pro}

While quenching is often discussed in the framework of the star-forming and quenched dichotomy, galaxies with similar star formation states can exhibit diverse structural properties \citep{fang2013, Barro2017ApJ, Wang2018a}. Such structural diversity may provide additional clues about the evolutionary pathways of massive galaxies. 

\begin{figure}
    \centering
    \includegraphics[width=1\linewidth]{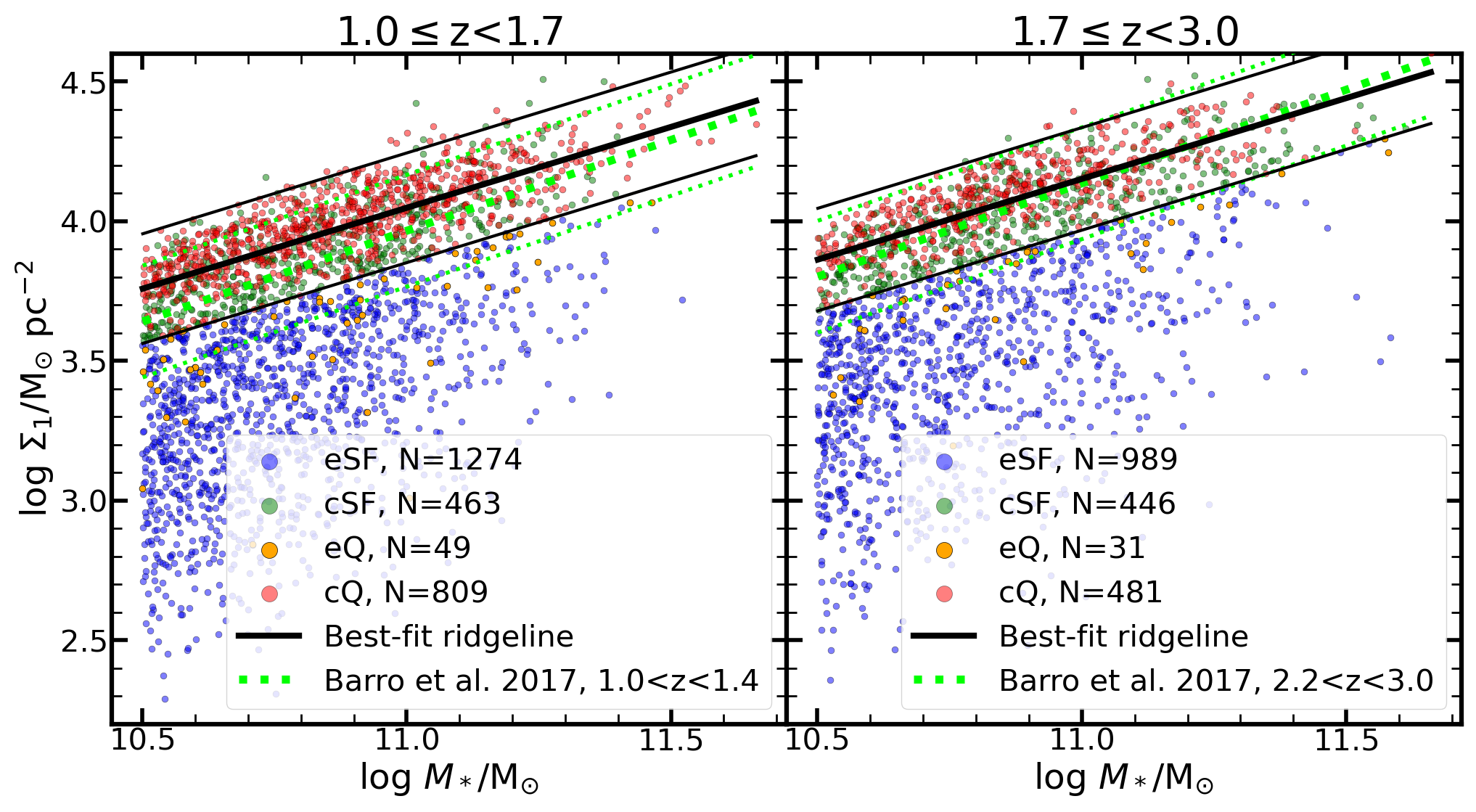}
    \caption{The \(\Sigma_1-M_\ast\) plane for different galaxy populations. The left and right panels show the galaxy distributions in the redshift bins \(1\leq z<1.7\) and \(1.7\leq z<3\), respectively. The blue, green, orange, and red points correspond to extended star-forming (eSF), compact star-forming (cSF), extended quenched (eQ), and compact quenched (cQ) galaxies, respectively. The thick black solid line indicates the best-fitting \(\Sigma_1-M_\ast\) ridgeline of quenched galaxies, while the thin black solid lines represent the \(\pm2\sigma\) boundaries used to define compact and extended populations. The lime dotted lines represent the fitted ridgeline and the \(\pm2\sigma\) scatter of the \(\Sigma_1-M_\ast\) relation for quenched galaxies from \cite{Barro2017ApJ}.}
    \label{fig_ridge}
\end{figure}

We adopt the method described in Section~\ref{sec_fit_ridge} to fit the quenched ridgeline and separate the galaxy sample into extended and compact subsamples. The quenched ridgeline and the corresponding scatter are listed in Table~\ref{tab_ridge} and also illustrated in Figure~\ref{fig_ridge}. The quenched galaxies in our COSMOS-Web sample follow a tight $\Sigma_1-M_\ast$ relation, with a scatter of $\sim0.1$ dex. The fitted ridgelines are in good agreement with those reported by \cite{Barro2017ApJ}, with comparable slopes ($\sim0.58$ vs. $\sim0.65$) and a similarly tight distribution around the relation ($\sim0.1$ dex scatter). 

\begin{table}[H]
    \centering
    \caption{Best-fitting quenched ridgeline parameters and scatters for the two redshift bins.}
    \setlength{\tabcolsep}{15.5pt}
    \begin{tabular}{c c c c}
    \hline
    \hline
    & a & b &  $\sigma_{\rm ridge}$ \\
    \hline
    $1\leq z<1.7$    & 4.05  & 0.58 & 0.1 \\
    $1.7\leq z < 3$  & 4.15  & 0.58 & 0.09  \\
    \hline
    \hline
    \end{tabular}
    \label{tab_ridge}
\end{table}

\begin{figure*}
    \centering    
    \includegraphics[width=1\linewidth]{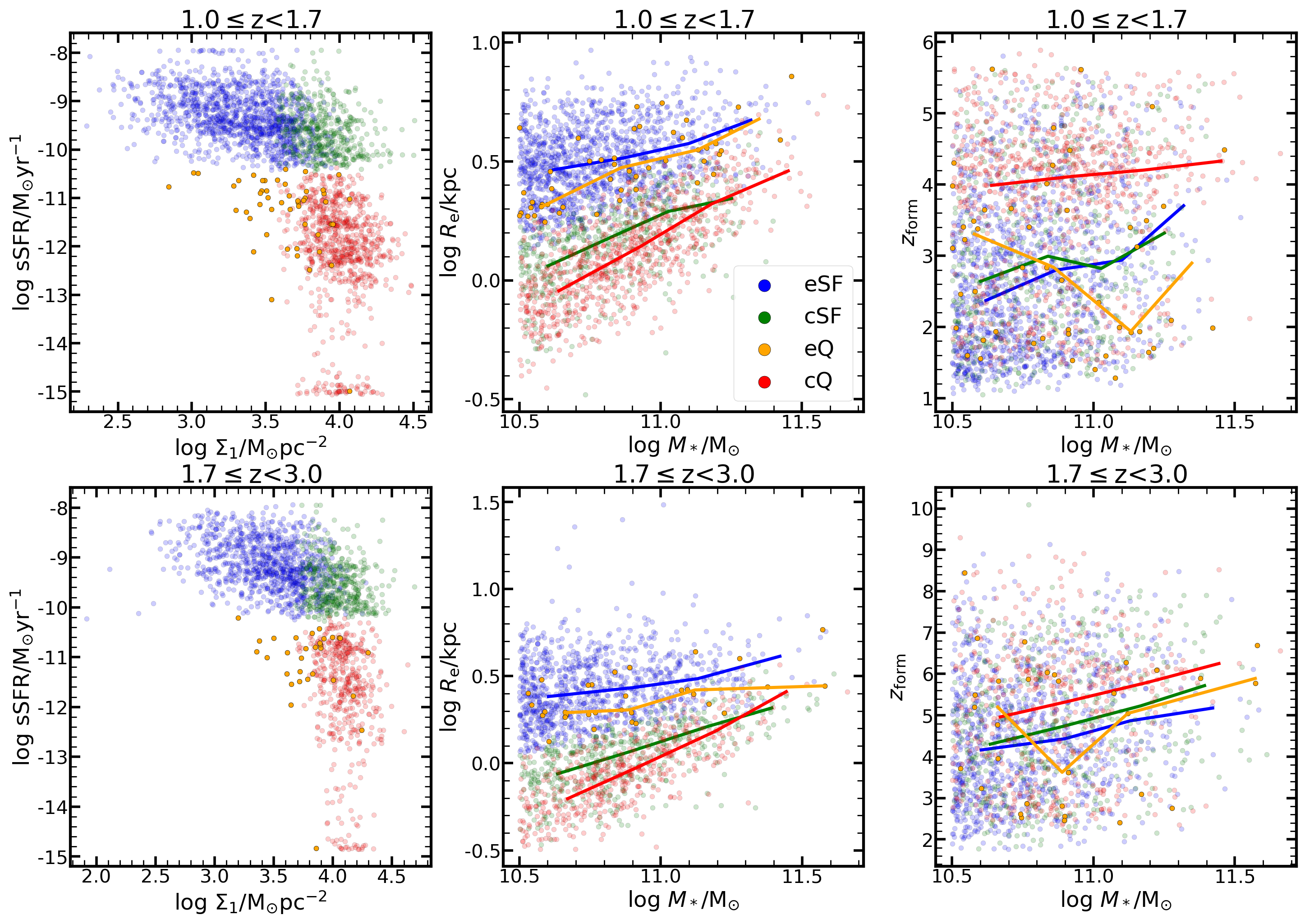}
    \caption{The distributions of eSF, cSF, eQ and cQ galaxies in different parameter spaces. The upper and lower rows show the galaxy distributions in the redshift bins \(1\leq z<1.7\) and \(1.7\leq z<3\), respectively. Blue, green, orange, and red symbols represent eSF, cSF, eQ, and cQ populations, respectively. The left, middle, and right columns correspond to the \(\Sigma_1\)--sSFR, $M_*$--$R_{\rm e}$, and $M_*$--$z_{\rm form}$ planes, respectively. The solid lines with different colors indicate the median relations of the corresponding galaxy populations.}
    \label{fig_para_dis}
\end{figure*}

We then compare the distributions of the extended star-forming (eSF), compact star-forming (cSF), extended quenched (eQ), and compact quenched (cQ) populations in different parameter spaces. The results are presented in Figure~\ref{fig_para_dis}.

Across both redshift bins, galaxies exhibit a clear L-shaped distribution in the sSFR--$\Sigma_1$ plane, consistent with previous studies (e.g., \citealt{fang2013, Barro2017ApJ, Wang2018a}) and highlighting the strong association between $\Sigma_1$ and star formation activity. Star-forming galaxies primarily occupy the high-sSFR regime and span a broad range of $\Sigma_1$, whereas quenched galaxies are concentrated at low sSFR and high $\Sigma_1$. Only a small fraction of quenched galaxies are found in the low-$\Sigma_1$ regime; in our sample, extended quenched galaxies account for only $\sim6\%$ of the quenched population (see Table~\ref{tab_samps}), with the majority of quenched galaxies classified as compact.

The separation between star-forming and quenched populations is also evident in their $M_*$--$R_{\rm e}$ and $M_*$--$z_{\rm form}$ relations. All populations exhibit positive correlations between $M_*$ and $R_{\rm e}$. At fixed $M_*$, quenched galaxies are generally more compact than star-forming galaxies (e.g., \citealt{vanderWel2014, Barro2017ApJ, Yang2025}), with smaller median $R_{\rm e}$ values across the stellar mass range. Extended galaxies have larger median $R_{\rm e}$ values than compact galaxies in both star-forming and quenched populations. Within each structural class, star-forming galaxies show slightly larger $R_{\rm e}$ values than quenched galaxies.

The right panels show $z_{\rm form}$ as a function of $M_*$. At fixed $M_*$, compact quenched galaxies exhibit higher median $z_{\rm form}$ values than star-forming galaxies in both redshift bins. The difference between the two populations is larger in the lower-redshift bin ($1\leq z<1.7$) than in the higher-redshift bin ($1.7\leq z<3$), suggesting a stronger separation in their inferred stellar formation histories at later cosmic times. Among star-forming galaxies, compact systems show slightly higher median $z_{\rm form}$ values than extended systems, although the difference is relatively small. The trend for extended quenched galaxies is less constrained due to their limited sample size, resulting in larger fluctuations. These variations in $z_{\rm form}$ provide additional constraints on the evolution of structurally distinct populations, which are further discussed in Section~\ref{sec_d1}.

The comparison of different structural populations further highlights the diversity of star-forming galaxies. A non-negligible fraction of star-forming galaxies reach high $\Sigma_1$ values comparable to the compact quenched galaxies \citep[e.g.,][]{Barro2017ApJ,Wang2018a}, demonstrating that high $\Sigma_1$ alone does not uniquely determine the quenching state. This motivates further investigation into whether structurally distinct populations are associated with different large-scale environments (see Section~\ref{sec_env}).

\subsection{The environmental dependence of structurally distinct galaxy populations}\label{sec_env}

\begin{figure*}
    \centering
    \includegraphics[width=1\linewidth]{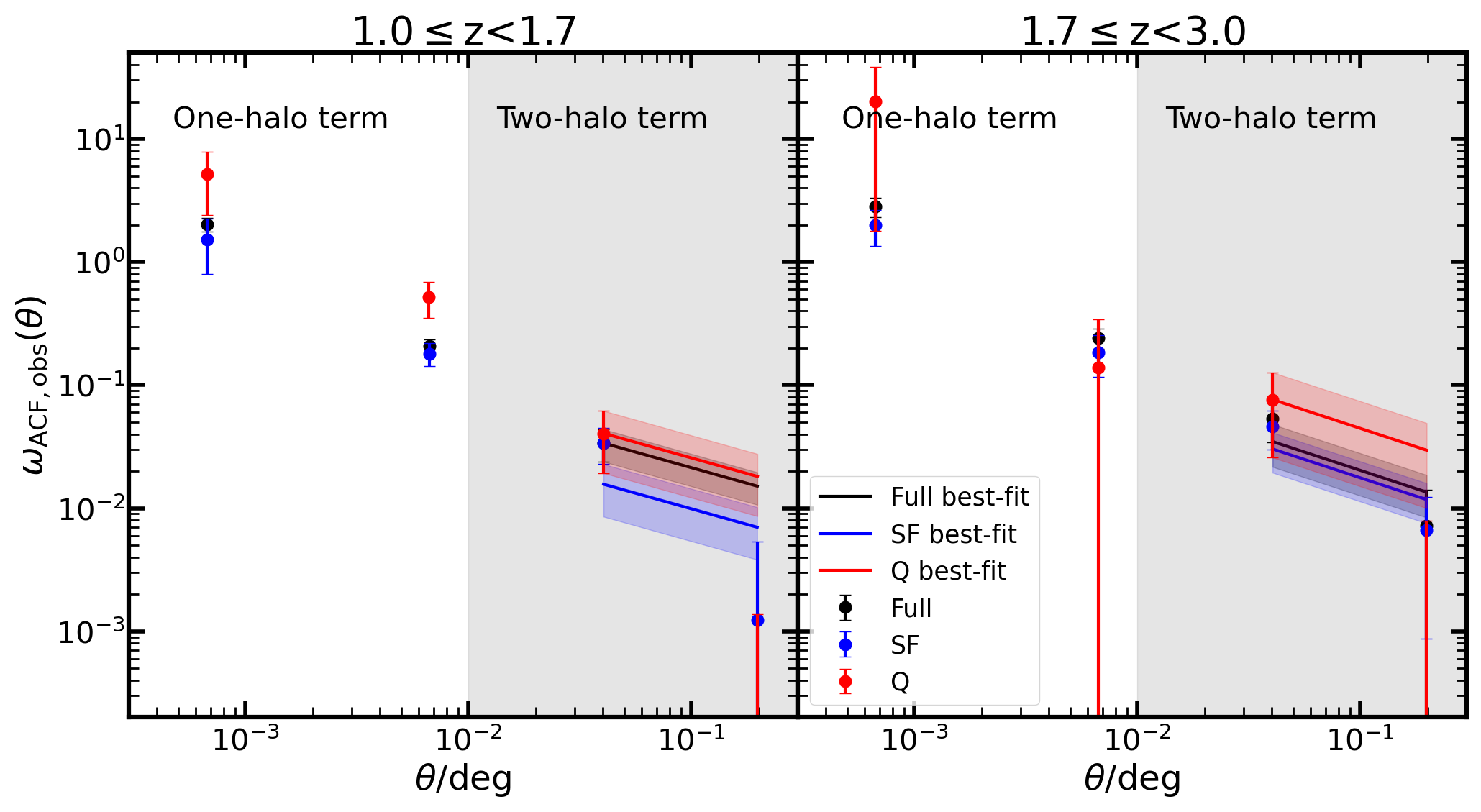}
    \caption{Observed angular auto-correlation functions of different galaxy samples. The left and right panels correspond to the redshift bins \(1\leq z<1.7\) and \(1.7\leq z<3\), respectively. Black, blue, and red symbols with error bars represent the ACFs of full, star-forming and quenched galaxies, respectively. The grey shaded regions indicate the approximate two-halo regime, while the unshaded regions denote the approximate one-halo regime. The solid lines show the best-fitting power-law models, with the associated uncertainties indicated by the shaded regions.}
    \label{fig_ori_corr0}
\end{figure*}

Galaxy clustering provides a direct probe of the connection between galaxy properties and their dark matter halos. Before investigating the environmental dependence of structurally distinct galaxy populations, we first examine whether the conventional star-forming and quenched classification is associated with different large-scale environments. Figure~\ref{fig_ori_corr0} and Table~\ref{tab_samps} present the measured clustering signals, galaxy biases, and inferred halo masses for the full, star-forming, and quenched samples in the two redshift bins, respectively. The methods used to estimate the galaxy biases and halo masses are described in Section~\ref{sec_bias_Mh}. The best-fit parameters of the power-law models used to characterize the angular correlation functions are summarized in Table~\ref{tab_fit} in the Appendix~\ref{sec_pl}. All clustering measurements are performed using stellar mass- and redshift-matched samples (see Section~\ref{sec_matching}).

At \(1\leq z<1.7\), the full, star-forming, and quenched populations show broadly similar large-scale clustering signals (\(\theta\gtrsim0.01\)). However, the inferred bias of quenched galaxies is higher than that of both the full and star-forming samples, with \(b_{\rm Q}=2.12^{+0.50}_{-0.66}\), compared with \(b_{\rm full}=1.81^{+0.25}_{-0.29}\) and \(b_{\rm SF}=1.20^{+0.25}_{-0.32}\). At \(1.7\leq z<3\), quenched galaxies show a similar tendency toward a higher inferred bias, with \(b_{\rm Q}=5.06^{+1.46}_{-2.11}\), \(b_{\rm full}=3.11^{+0.54}_{-0.66}\), and \(b_{\rm SF}=2.84^{+0.47}_{-0.57}\). However, the larger uncertainties in the inferred biases, mainly caused by the smaller number of quenched galaxies, prevent a statistically significant distinction among these populations. Overall, the conventional star-forming/quenched classification reveals a marginal difference in large-scale clustering for massive galaxies at \(1<z<3\). Although not statistically significant given the current uncertainties, the systematically higher bias of quenched galaxies suggests a possible tendency for quenched galaxies to reside in more massive halos \citep{Li06a, Zhang2021}. 

\begin{figure*}
    \centering
    \includegraphics[width=1\linewidth]{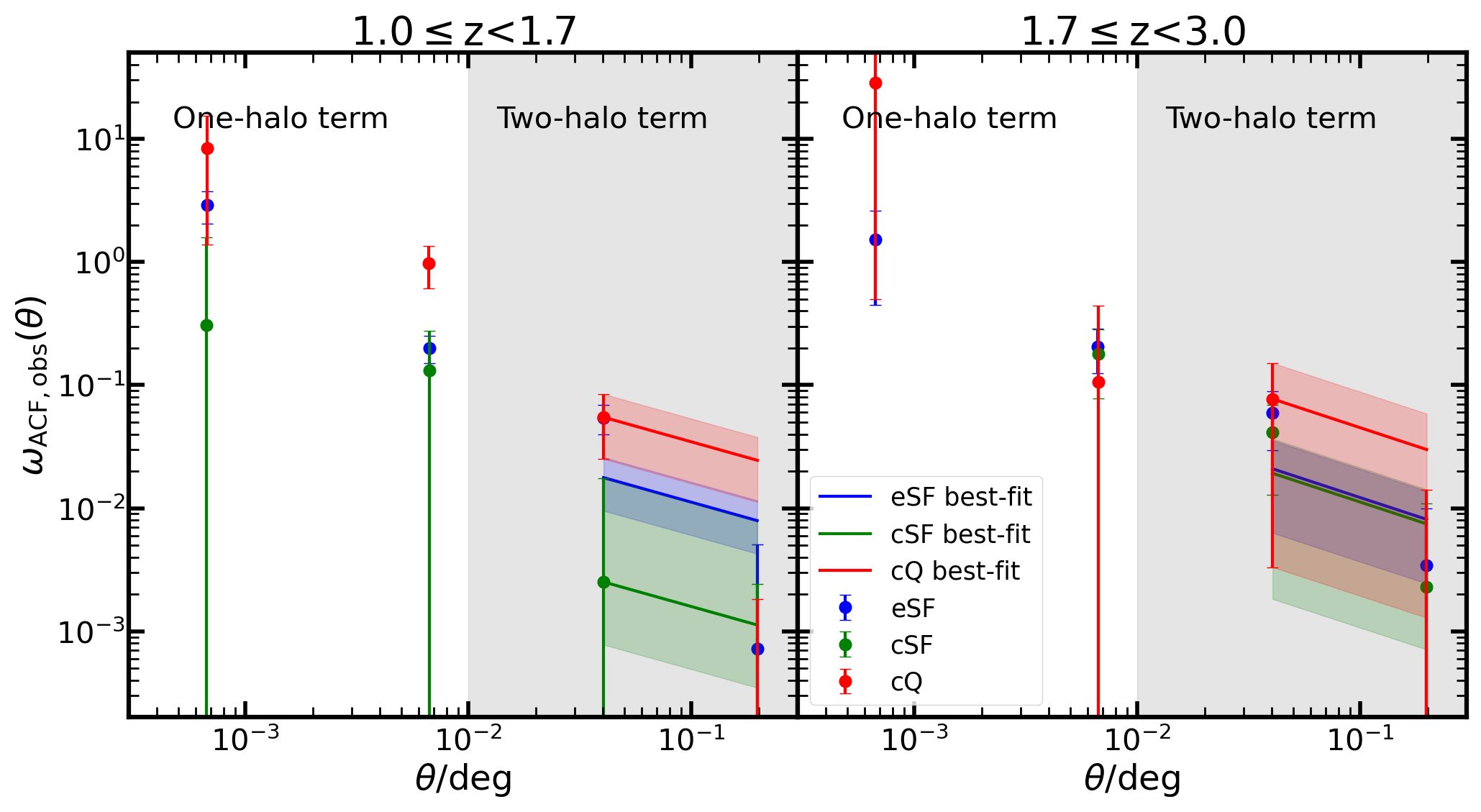}
    \caption{Similar to Figure~\ref{fig_ori_corr0}, but for the extended star-forming (eSF, blue), compact star-forming (cSF, green) and compact quenched (cQ, red) galaxy samples.}
    \label{fig_ori_corr}
\end{figure*}

We further examine whether structural properties provide additional information on the galaxy--halo connection. Figure~\ref{fig_ori_corr} presents clustering measurements of eSF, cSF, and cQ galaxies, with the eQ sample omitted due to limited sample size and large clustering uncertainties. Table~\ref{tab_samps} lists their galaxy biases and inferred halo masses, and Table~\ref{tab_fit} in Appendix~\ref{sec_pl} summarizes the best-fit power-law model parameters.

At \(1\leq z<1.7\), the structural populations show a tentative difference in their clustering amplitudes, with cQ galaxies exhibiting the highest inferred bias among the three populations. The inferred bias of cQ galaxies is \(b_{\rm cQ}=1.93^{+0.47}_{-0.62}\), compared with \(b_{\rm eSF}=1.02^{+0.21}_{-0.27}\) and \(b_{\rm cSF}=0.36^{+0.59}_{-0.16}\). However, the uncertainties, particularly for the cSF population, prevent a statistically significant separation among these populations. The comparison between eSF and cSF galaxies shows no significant difference in their large-scale clustering properties within the current uncertainties.

At \(1.7\leq z<3\), the clustering differences among structural populations become less significant. The clustering signals of eSF, cSF, and cQ galaxies overlap substantially within the uncertainties, with inferred biases of \(b_{\rm cQ}=4.78^{+1.91}_{-3.79}\), \(b_{\rm cSF}=2.00^{+0.76}_{-1.38}\), and \(b_{\rm eSF}=2.20^{+0.67}_{-0.99}\). The higher central value of the cQ bias suggests a possible enhancement of clustering for cQ galaxies, whereas eSF and cSF galaxies show similar clustering properties.

Taken together, structural classification provides additional information on the galaxy--halo connection beyond the conventional star-forming/quenched division. We find no significant difference in large-scale clustering between structurally distinct star-forming galaxies, while cQ galaxies show a tentative tendency toward stronger clustering and higher inferred halo masses. Previous clustering studies of cSF and cQ galaxies at $z\sim1.2-3$ have reported comparable clustering strengths for these two populations (e.g., \citealt{Lin2017}), consistent with our finding that the difference between cSF and cQ galaxies is not statistically significant within the current uncertainties, although cQ galaxies exhibit higher central values of the inferred bias and halo mass. In contrast, eSF and cSF galaxies show similar large-scale clustering signals despite their different structural properties, consistent with previous findings that central density-related properties are more closely associated with quenching than with environment \citep{Wang2018a}.

\section{Discussion}\label{sec_discussion}

\subsection{$\Sigma_1$ as a key predictor of quenching at cosmic noon}\label{sec_d1}

Our analysis shows that central stellar density, $\Sigma_1$, is the strongest predictor of galaxy quenching during cosmic noon. Among the galaxy properties considered in the Random Forest analysis, $\Sigma_1$ exhibits the highest predictive importance, substantially exceeding that of stellar mass. This indicates that the internal distribution of stellar mass provides a closer observational connection to quenching than global stellar mass for massive galaxies at $1\leq z<3$. 

Previous studies across different redshift ranges have established a strong association between central stellar density and quenching, often identifying characteristic $\Sigma_1$ thresholds separating star-forming and quenched populations \citep{Cheung2012, fang2013, Barro2017ApJ, Mosleh2017, Whitaker2017}. These studies have demonstrated the importance of central density as a quenching-related structural parameter, but mainly focused on correlations or structural differences between quenched and star-forming galaxies. The relative importance of $\Sigma_1$ compared with other quenching-related properties has been less explored. Recent machine-learning studies have enabled quantitative comparisons among different galaxy properties, showing that central galaxy properties, including central velocity dispersion and bulge-related quantities, provide stronger predictions of quenching than global stellar mass or environmental indicators \citep{Bluck2020b, Bluck2020, Bluck2022}. Extending this approach to cosmic noon, our analysis simultaneously evaluates structural, stellar population, and environmental properties, revealing the dominant predictive role of $\Sigma_1$ for quenching at $1\leq z<3$.

The strong association between $\Sigma_1$ and quenching is further supported by the quenched fraction distribution. In our sample, the quenched fraction increases rapidly above $\log\Sigma_1\simeq3.6-3.7$, while showing a much weaker dependence on stellar mass. This rapid increase suggests that galaxies above this central density regime are preferentially associated with the quenched population. At fixed stellar mass, galaxies with higher $\Sigma_1$ exhibit more centrally concentrated stellar distributions and generally more prominent bulge components \citep{fang2013, Lang2014}.

Several mechanisms may contribute to the buildup of central stellar components. Gas-rich processes, including cold gas accretion, gravitational instabilities, and gas-rich interactions, can drive baryons toward galaxy centers, enhancing central gas concentration and subsequent star formation, thereby increasing stellar concentration \citep{Dekel2014, Zolotov2015, Lapiner2023, Lyu2025, Maql2026}. Such dissipative compaction can produce cSF systems, or ``blue nuggets'', which have been proposed as an intermediate evolutionary stage between eSF and cQ galaxies \citep{Zolotov2015, Barro2017ApJ, Wang2018a}.

Our results are broadly consistent with this structural transformation scenario. We find that cSF galaxies have slightly higher $z_{\rm form}$ than eSF galaxies, but lower $z_{\rm form}$ than cQ galaxies. This ordering suggests that cSF galaxies may represent an intermediate population between eSF and cQ systems. However, this interpretation should be treated with caution. Differences in structure and formation history among eSF, cSF, and cQ galaxies do not necessarily imply a direct evolutionary sequence, as progenitor bias may contribute to the observed population differences. Galaxies that quenched at different epochs may enter the quenched population with different structural properties and formation histories \citep{Lilly2016, Tacchella2017, Yang2025}. With the current data, we cannot distinguish whether these differences primarily reflect genuine evolutionary transformations, progenitor bias, or a combination of both.

In addition to internal galaxy properties, we also examined the role of environment. The relatively weak contribution of local overdensity $(1+\delta)$ suggests that environment provides less predictive information than internal structural properties for our massive galaxies \citep{Wang2018a}. This does not imply that halo environments are irrelevant, as halo-related processes may regulate gas accretion, thereby influencing quenching \citep{Dekel2014, Woo2015}. We next discuss how halo environments are connected to different structural populations, from eSF to cQ galaxies.

\subsection{The evolutionary connection between structural transformation, quenching, and halo environments}

The role of halo environments in structural transformation and quenching has been explored in previous studies, including comparisons of the clustering signals between cSF and cQ galaxies \citep{Lin2017}. However, whether the full structural diversity of eSF, cSF, and cQ galaxies is accompanied by systematic differences in their halo environments remains less well constrained. Using a homogeneous JWST COSMOS-Web sample, our clustering measurements allow us to examine the galaxy--halo connection across these structural populations.

A key result from our clustering analysis is that eSF and cSF galaxies exhibit comparable galaxy biases within the uncertainties. Despite their substantial differences in $\Sigma_1$ and size, these two populations appear to reside in halos with similar masses. If structural differences between eSF and cSF galaxies were primarily driven by different halo environments, we would expect a measurable difference in their large-scale bias. The absence of such a difference suggests that halo mass variations alone cannot account for the observed structural transformation, pointing to an important role of internal baryonic processes in shaping galaxy structures \citep{Dekel2014, Zolotov2015, Lapiner2023}. The lack of significant halo mass differences may also suggests that structural transformation can occur on timescales shorter than major evolution in halo mass.

In contrast, cQ galaxies show a tendency toward higher inferred halo masses than cSF galaxies. This difference may suggests that, after quenching, stellar masses evolve relatively slowly while host halos continue to grow, allowing older quenched populations to reside in progressively more massive halos \citep{Zhang2021}. This interpretation is consistent with previous studies showing that quenching is associated with the combination of dense central structures and sufficiently massive halos \citep{Woo2015, Guo2021}. In this picture, the buildup of $\Sigma_1$ provides an important structural condition for quenching, while subsequent halo growth shapes the evolution of the quenched population.

Taken together, our results favor a picture in which internal structural evolution and halo growth need not proceed synchronously. The substantial structural differences between eSF and cSF can develop within broadly similar halo environments, structural evolution can occurr on a shorter timescale than substantial halo growth. Meanwhile, the more massive halos associated with cQ may reflect continued halo growth after or around the onset of quenching.

\subsection{Caveats and future prospects}

Our analysis provides observational constraints on the galaxy--halo connection during cosmic noon, but several limitations remain. First, the limited area and sample size of the COSMOS-Web field introduce significant uncertainties in clustering measurements. Upcoming wide-field surveys such as Euclid, Roman, CSST, and LSST will be essential to establish these trends with improved statistical significance.

Second, our structural measurements are based on Sérsic profile fitting, with $\Sigma_1$ inferred from the stellar mass enclosed within 1 kpc. This model-dependent approach may introduce uncertainties due to assumptions about the underlying light profile, PSF correction, and spatial variations in stellar populations \citep{Maql2026}. Future analyses based on spatially resolved stellar mass maps will enable more direct and robust measurements of $\Sigma_1$.

Finally, although $\Sigma_1$ shows the strongest association with quenching in our analysis, this correlation alone does not establish a causal connection \citep{Lilly2016}. Future observations of molecular gas reservoirs, gas kinematics, and black hole activity will be required to identify the physical mechanisms linking structural transformation to the suppression of star formation. 

\section{Summary}\label{sec_summary}
In this work, we investigate the connection between galaxy structural properties, star formation activity, and large-scale environments for massive galaxies at $1<z<3$ using JWST/COSMOS-Web observations. By combining structural measurements, Random Forest analysis, and galaxy clustering measurements, we quantify the role of different galaxy properties in predicting quenching and explore the halo environments of structurally distinct galaxy populations. Our main conclusions are summarized as follows:

\begin{enumerate}

\item We quantify the relative importance of different galaxy properties in predicting galaxy quenching, including $M_*$, $\Sigma_1$ $R_{\rm e}$, $z_{\rm form}$, and ($1+\delta$). Among these parameters, $\Sigma_1$ provides the strongest predictive power, while the additional parameters contribute only marginal improvements. The strong dependence of quenched fraction on $\Sigma_1$ demonstrates that the buildup of central stellar mass surface density is closely associated with quenching in massive galaxies during cosmic noon.

\item We systematically investigate the large-scale environments of structurally classified galaxy populations, including extended star-forming (eSF), compact star-forming (cSF), and compact quenched (cQ) galaxies. We find no significant difference in the clustering properties of eSF and cSF galaxies despite their distinct structures, suggesting similar halo environments for these two star-forming populations. cQ galaxies exhibit higher central values of clustering amplitude and inferred halo mass, although the uncertainties prevent a statistically significant distinction from the other populations. The clustering of extended quenched galaxies remains unconstrained due to the limited sample size.

\end{enumerate}

These results emphasize the role of central stellar mass surface density in galaxy quenching and provide new constraints on the halo connection of structurally distinct populations during cosmic noon. Future wide-field surveys will extend this analysis with larger samples and improved statistical precision.

\section{Acknowledgements}

This work is supported by the China Postdoctoral Science Foundation-Anhui Joint Support Program under Grant Number 2025T004AH and Anhui Postdoctoral Scientific Research Program Foundation (No. BH2030000803). This work is supported by the National Natural Science Foundation of China (NSFC, Nos. 12595312, 12192224 ). HYW thanks the support of CAS Project for Young Scientists in Basic Research, Grant No. YSBR-062, the New Cornerstone Science Foundation through the XPLORER PRIZE, and the China Manned Space Program
with grant no. CMS-CSST-2025-A04.  The work is also supported by the Supercomputer Center of University of Science and Technology of China.

\bibliography{ref}{}
\bibliographystyle{aasjournal}

\begin{appendix}

\section{Robustness tests with alternative sample selection}
\label{sec_alter}

\FloatBarrier

To examine the robustness of our results against the sample selection criteria, we construct alternative star-forming and quenched samples by including the galaxies with inconsistent color and sSFR classifications that were excluded from the fiducial samples. The corresponding sample properties, including the number of galaxies, galaxy biases, and inferred halo masses, are summarized in Table~\ref{tab_s_samps}.

\begin{figure}
    \centering
    \includegraphics[width=0.8\linewidth]{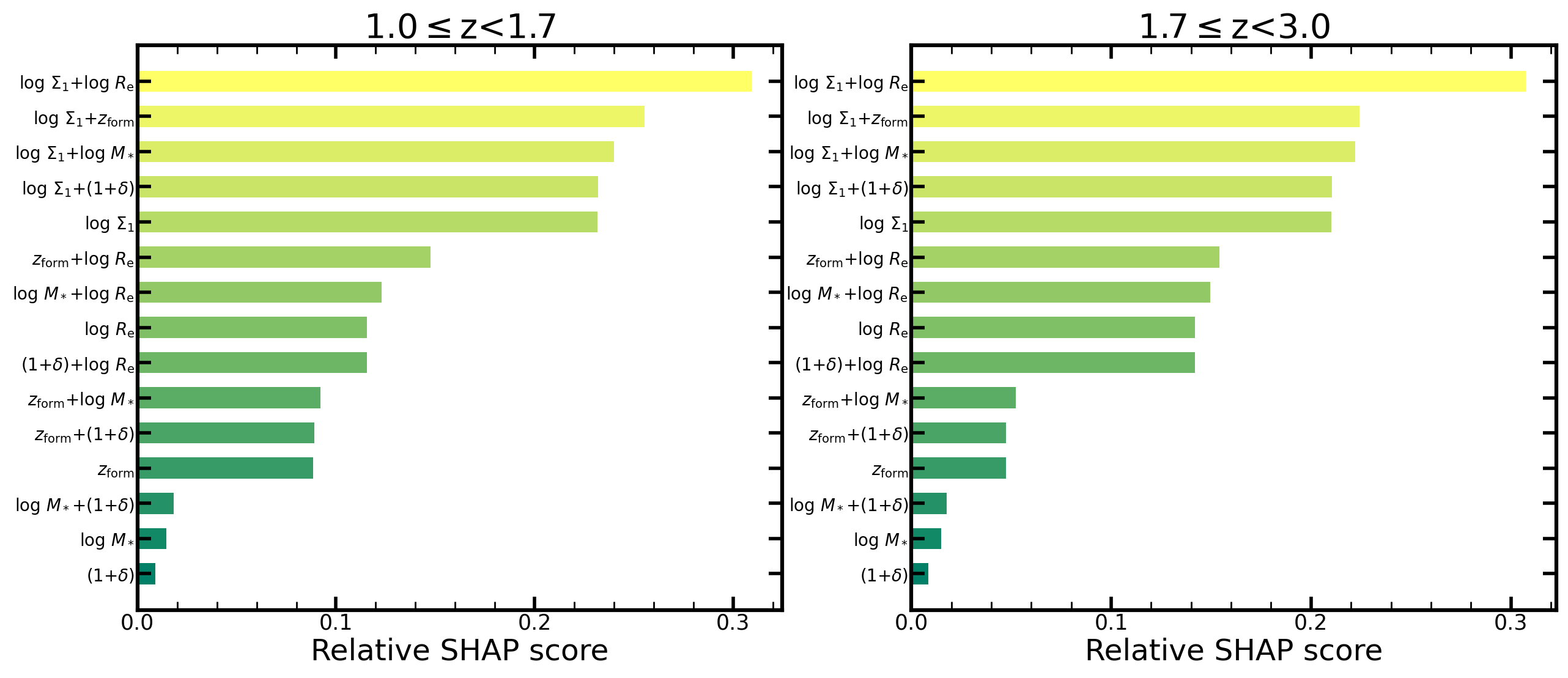}
    \caption{Similar to Figure~\ref{fig_SHAP}, but for the alternative galaxy sample selection.}
    \label{fig_s_shap}
\end{figure}

The SHAP importance of different galaxy properties for the alternative sample is presented in Figure~\ref{fig_s_shap}. The relative importance of galaxy properties in predicting quenching remains unchanged: $\Sigma_1$ is the most important predictor of galaxy quenching, followed by $R_{\rm e}$, $z_{\rm form}$, $M_*$, and $(1+\delta)$.

\begin{figure}
    \centering
    \includegraphics[width=0.6\linewidth]{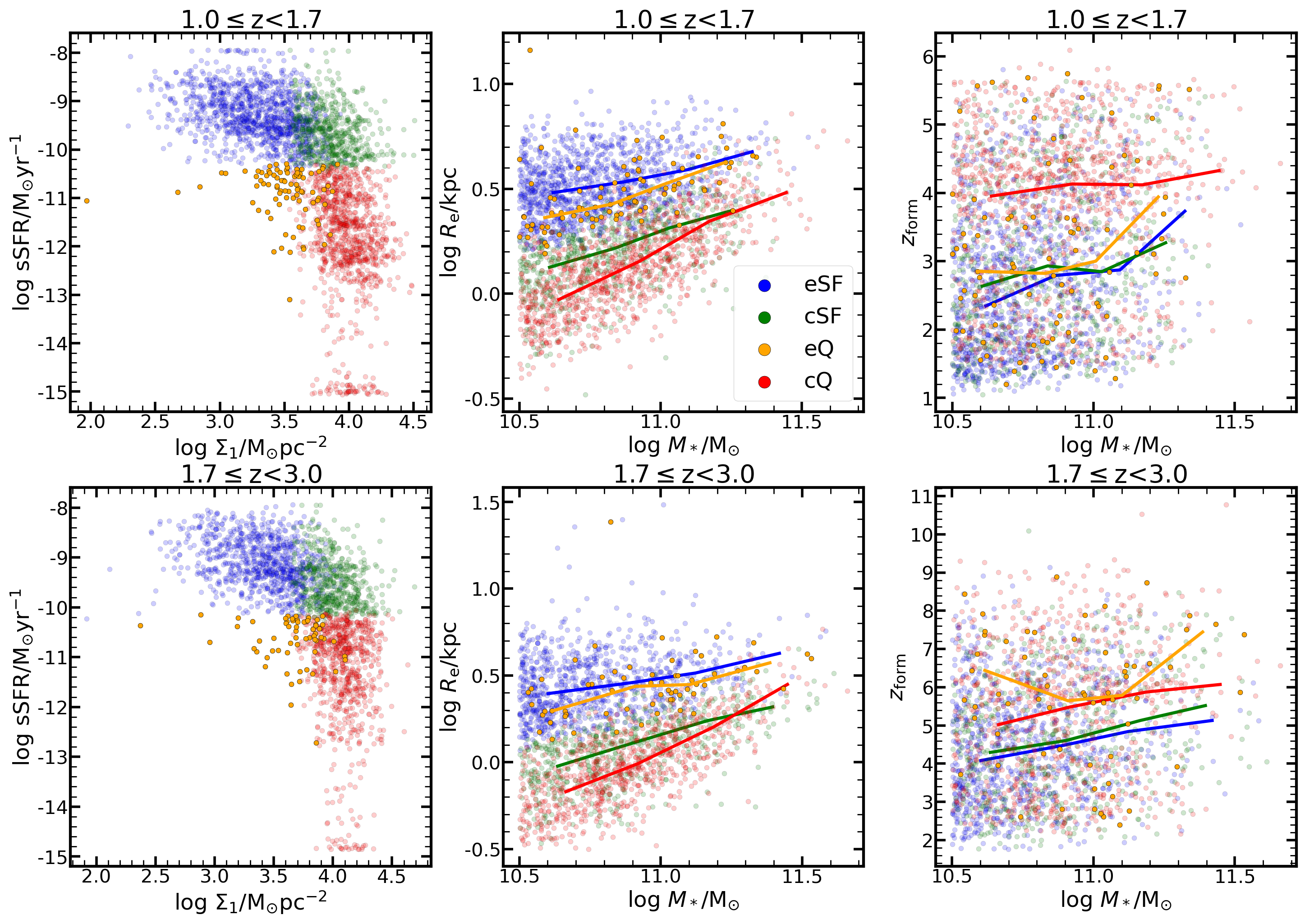}
    \caption{Similar to Figure~\ref{fig_para_dis}, but for the alternative galaxy sample selection.}
    \label{fig_s_para_dis}
\end{figure}

The galaxy distributions obtained with the alternative selection are shown in Figure~\ref{fig_s_para_dis}. The distributions of galaxy properties, including structural parameters, sSFR and $z_{\rm form}$, among extended and compact star-forming and quenched populations remain broadly consistent with those of the fiducial sample (Figure~\ref{fig_para_dis}).

\begin{figure}
    \centering
    \includegraphics[width=0.8\linewidth]{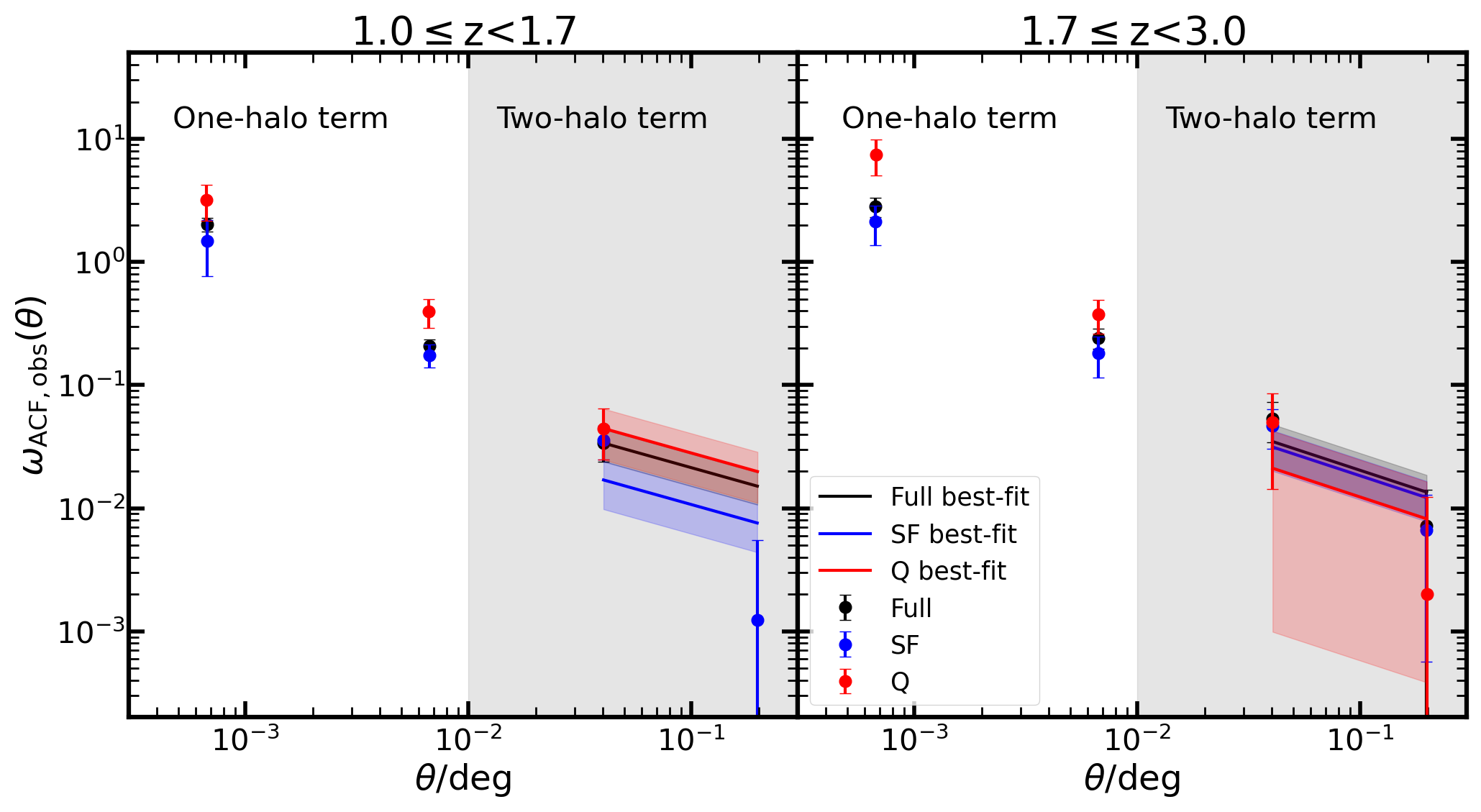}
    \caption{Similar to Figure~\ref{fig_ori_corr0}, but for the alternative galaxy sample selection. The clustering measurements of the full, star-forming, and quenched populations are shown.}
    \label{fig_s_corr0}
\end{figure}

The clustering measurements of the full, star-forming, and quenched populations are shown in Figure~\ref{fig_s_corr0}. At \(1\leq z<1.7\), quenched galaxies exhibit a higher inferred bias than star-forming galaxies. At \(1.7\leq z<3\), however, the inferred biases of the three populations overlap within the uncertainties, mainly due to the large uncertainty of the quenched sample.

\begin{figure}[H]
    \centering
    \includegraphics[width=0.8\linewidth]{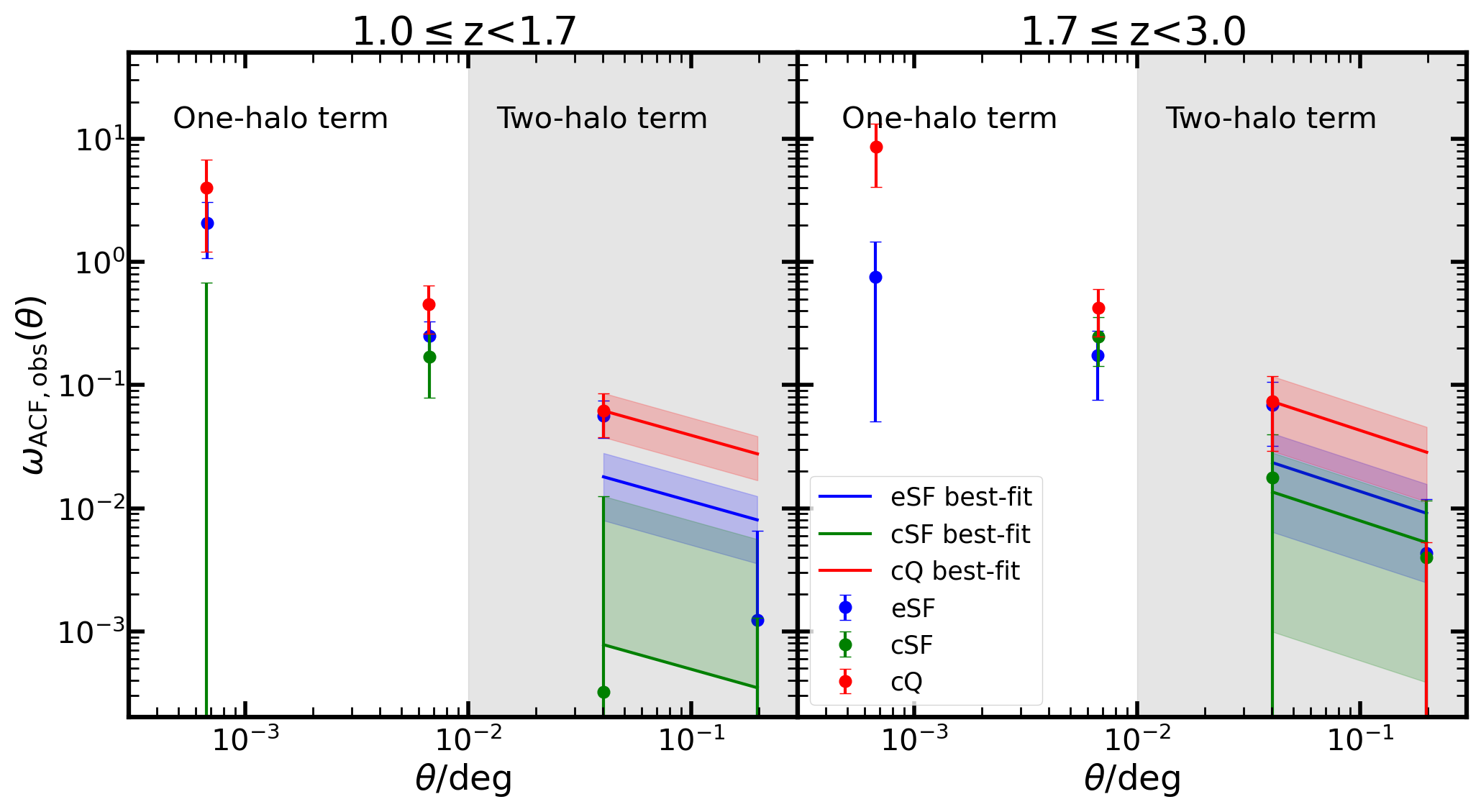}
    \caption{Similar to Figure~\ref{fig_ori_corr}, but for the alternative galaxy sample selection. The clustering measurements of the structurally classified populations are shown.}
    \label{fig_s_corr}
\end{figure}

Figure~\ref{fig_s_corr} presents the clustering measurements of the structurally selected populations. The alternative selection yields the same qualitative trend as the fiducial analysis: cQ galaxies show a tendency toward stronger clustering, while cSF and eSF galaxies exhibit comparable clustering amplitudes within the uncertainties.

\begin{table}[t]
\centering

\caption{Number of galaxies, galaxy biases, and inferred halo masses of different galaxy subsamples selected using the alternative criteria.}
\label{tab_s_samps}

\begin{tabular*}{\linewidth}{@{\extracolsep{\fill}}l|ccc|ccc}

\hline
\hline

Subsample
& \multicolumn{3}{c|}{$1\leq z<1.7$}
& \multicolumn{3}{c}{$1.7\leq z<3$}
\\

&
$N_{\rm gal}$
& $b_{\rm g}$
& $\log(M_{\rm h}/M_{\odot})$
&
$N_{\rm gal}$
& $b_{\rm g}$
& $\log(M_{\rm h}/M_{\odot})$
\\

\hline

Star-forming (SF)
& 1752
& $1.25^{+0.24}_{-0.30}$
& $11.58^{+0.43}_{-0.92}$
& 1450
& $2.90^{+0.48}_{-0.58}$
& $12.36^{+0.24}_{-0.38}$
\\

Quenched (Q)
& 1089
& $2.15^{+0.43}_{-0.55}$
& $12.71^{+0.28}_{-0.54}$
& 854
& $2.53^{+1.05}_{-1.98}$
& $12.14^{+0.54}_{-2.98}$
\\

Extended star-forming (eSF)
& 1160
& $1.04^{+0.26}_{-0.35}$
& $10.90^{+0.68}_{-1.74}$
& 897
& $2.38^{+0.75}_{-1.14}$
& $11.98^{+0.46}_{-1.54}$
\\

Compact star-forming (cSF)
& 592
& $0.20^{+0.61}_{-0.00}$
& $9.17^{+0.54}_{-0.00}$
& 553
& $1.68^{+0.74}_{-1.22}$
& $11.25^{+0.75}_{-2.08}$
\\

Extended quenched (eQ)
& 87
& --
& --
& 62
& --
& --
\\

Compact quenched (cQ)
& 1002
& $1.99^{+0.36}_{-0.44}$
& $12.50^{+0.27}_{-0.48}$
& 792
& $4.42^{+1.18}_{-1.64}$
& $12.93^{+0.30}_{-0.68}$
\\

\hline
\hline

\end{tabular*}
\end{table}

Overall, the alternative sample selection does not change the main conclusions of this work. The strong association between $\Sigma_1$ and quenching, as well as the slightly higher galaxy bias of quenched galaxies compared with other populations, remain robust against reasonable variations in sample selection.

\section{Stellar mass and redshift matching}\label{sec_weighting}

\begin{figure}[H]
    \centering
    \includegraphics[width=0.6\linewidth]{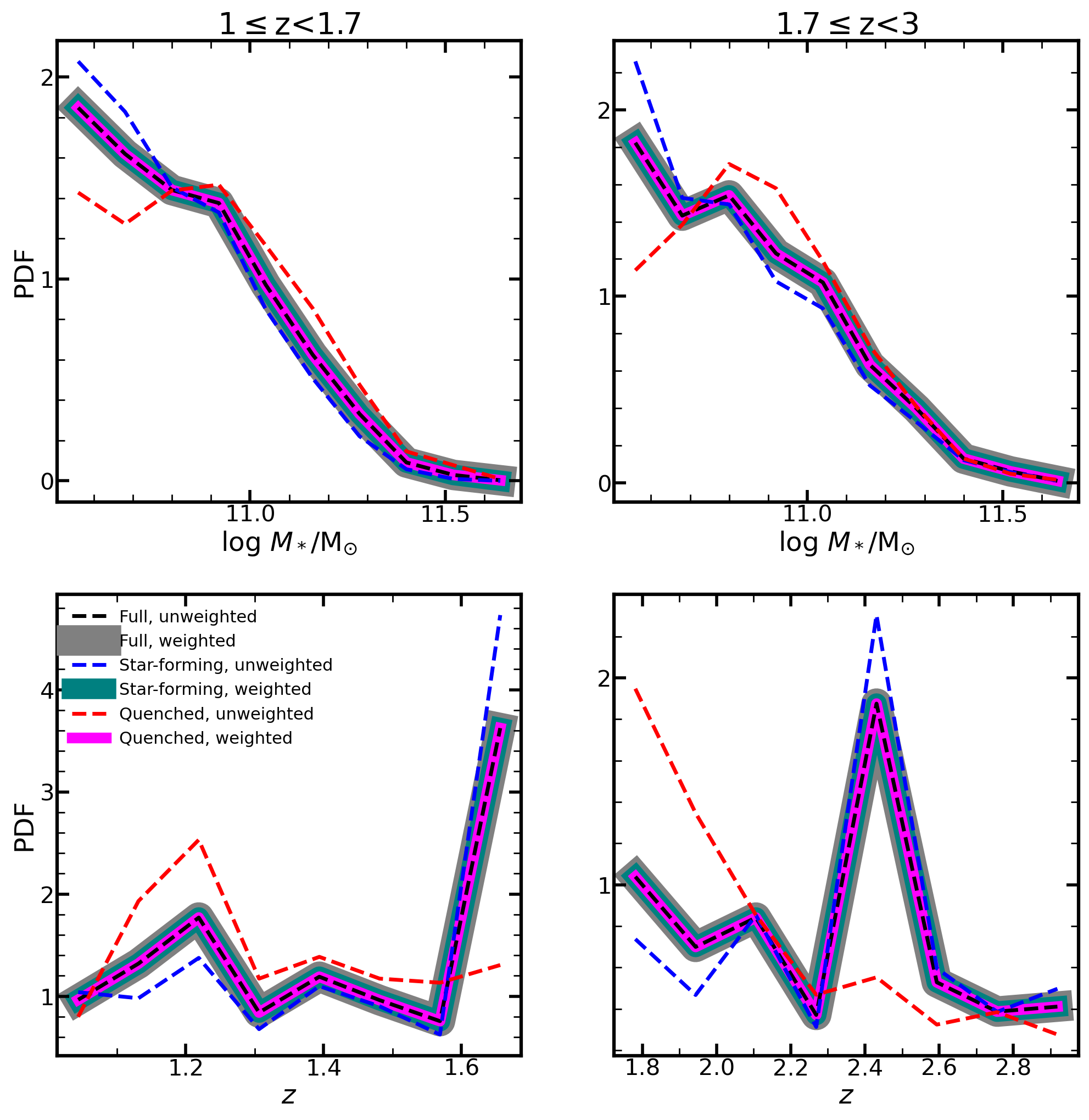}
    \caption{Stellar mass (upper row) and redshift (lower row) distributions of the galaxy samples before and after the iterative reweighting procedure. The left and right columns correspond to the redshift bins \(1\leq z<1.7\) and \(1.7\leq z<3\), respectively. The full, star-forming, and quenched populations are shown in black/grey, blue/teal, and red/magenta. Thick solid and thin dashed lines represent the weighted and the original unweighted distributions, respectively.}
    \label{fig_dis1}
\end{figure}

Figure~\ref{fig_dis1} shows the stellar mass and redshift distributions before and after the iterative reweighting procedure.

\section{Power-law fitting of angular correlation functions}\label{sec_pl}

The best-fit parameters of the power-law models used to describe the angular correlation functions are summarized in Table~\ref{tab_acf_fit}.

\begin{table}\label{tab_fit}
    \centering
    \small
    \setlength{\tabcolsep}{3pt}
    \caption{Power-law fitting results of the angular correlation functions for different galaxy samples.
    The amplitude $A_{\omega}$ is in units of $10^{-3}\rm\,\theta^{\gamma-1}$, and the correlation length $r_0$ is in units of cMpc/$h$.}
    \label{tab_acf_fit}
    \begin{tabular*}{\linewidth}{@{\extracolsep{\fill}}cccccc}
    \hline
    \hline
    $z$-bin & Sample & $\gamma$ & $C$ & $A_{\omega}$($\times 10^{-3}\theta^{\gamma-1}$) & $r_0$ (cMpc/$h$)\\
    \hline
    
    & Parent & 1.77 & 4.03 & $2.27^{+1.31}_{-1.31}$ & --\\
    & Full & 1.77 & 4.03 
    & $4.34^{+1.28}_{-1.28}$ 
    & $4.27^{+0.67}_{-0.76}$\\
    & SF & 1.77 & 4.03 
    & $2.01^{+0.91}_{-0.91}$ 
    & $2.70^{+0.64}_{-0.78}$\\
$1\leq z<1.7$
    & Q & 1.77 & 4.03 
    & $5.21^{+2.73}_{-2.73}$ 
    & $5.11^{+1.38}_{-1.75}$\\
    & eSF & 1.77 & 4.03 
    & $2.27^{+1.05}_{-1.05}$ 
    & $2.15^{+0.51}_{-0.64}$\\
    & cSF & 1.77 & 4.03 
    & $0.32^{+1.93}_{-0.22}$ 
    & $0.67^{+1.34}_{-0.33}$\\
    & cQ & 1.77 & 4.03 
    & $7.04^{+3.81}_{-3.81}$ 
    & $4.45^{+1.23}_{-1.59}$\\
    
    \hline
    
    & Parent & 1.83 & 4.57  & $1.74^{+1.11}_{-1.11}$ & --\\
    & Full & 1.83 & 4.57 
    & $3.51^{+1.33}_{-1.33}$ 
    & $5.59^{+1.07}_{-1.27}$\\
    & SF & 1.83 & 4.57 
    & $3.05^{+1.09}_{-1.09}$ 
    & $5.05^{+0.92}_{-1.09}$\\
$1.7\leq z<3$
    & Q & 1.83 & 4.57 
    & $7.70^{+5.09}_{-5.09}$ 
    & $9.52^{+3.04}_{-4.25}$\\
    & eSF & 1.83 & 4.57 
    & $2.11^{+1.47}_{-1.47}$ 
    & $3.75^{+1.26}_{-1.80}$\\
    & cSF & 1.83 & 4.57 
    & $1.94^{+1.75}_{-1.75}$ 
    & $3.36^{+1.42}_{-2.43}$\\
    & cQ & 1.83 & 4.57 
    & $7.79^{+7.46}_{-7.46}$ 
    & $8.78^{+3.88}_{-7.20}$\\

    \hline
    \hline
    \end{tabular*}
\end{table}

\end{appendix}
\end{document}